\documentclass[apj]{emulateapj}
\usepackage{multirow}
\usepackage[english]{babel}
\usepackage{natbib}
\usepackage{color}

\shortauthors{RAICHOOR ET AL.}
\shorttitle{EARLY-TYPE GALAXIES AT Z $\sim$ 1.3}
\slugcomment{ApJ - Accepted (18/02/2011)}

\usepackage{hyperref}

\begin{document}

\title{Early-type galaxies at z $\sim$ 1.3. II. Masses and ages of early-type galaxies in different environments and their dependence on stellar population model assumptions}
\author{
A. Raichoor\altaffilmark{1,2}, 
S.  Mei\altaffilmark{1,3},
F. Nakata\altaffilmark{4},
S.A.Stanford\altaffilmark{5,6,$\star$},
B.P. Holden\altaffilmark{7},
A. Rettura\altaffilmark{5,8},
M. Huertas-Company\altaffilmark{1,3},
M. Postman\altaffilmark{9},
P. Rosati\altaffilmark{10},
J.P. Blakeslee\altaffilmark{11},
R. Demarco\altaffilmark{12},
P. Eisenhardt\altaffilmark{13},
G. Illingworth\altaffilmark{7},
M.J. Jee\altaffilmark{5},
T. Kodama\altaffilmark{4,14},
M. Tanaka\altaffilmark{15},
R.L. White\altaffilmark{9}}

\altaffiltext{1}{GEPI, Observatoire de Paris, Section de Meudon, 5 Place J. Janssen, 92190 Meudon Cedex, France} 
\altaffiltext{2}{\textit{Current address}: Osservatorio Astronomico di Brera, via Brera 28, 20121 Milan, Italy; e-mail: \texttt{anand.raichoor@brera.inaf.it}} 
\altaffiltext{3}{Universit\'{e} Paris Denis Diderot, 75205 Paris Cedex 13, France}
\altaffiltext{4}{Subaru Telescope, National Astronomical Observatory of Japan, 650 North A�ohoku Place, Hilo, HI 96720, USA}
\altaffiltext{5}{Department of Physics, University of California, Davis, CA 95616, USA}
\altaffiltext{6}{Institute of Geophysics and Planetary Physics, Lawrence Livermore National Laboratory, Livermore, CA 94551, USA}
\altaffiltext{7}{ UCO/Lick Observatories, University of California, Santa Cruz 95065, USA}
\altaffiltext{8}{Department of Physics and Astronomy, Johns Hopkins University, Baltimore, MD 21218, USA}
\altaffiltext{9}{Space Telescope Science Institute, Baltimore, MD, USA}
\altaffiltext{10}{European South Observatory, Karl-Schwarzschild -Str. 2, D-85748, Garching, Germany}
\altaffiltext{11}{Herzberg Institute of Astrophysics, National Research Council of Canada, Victoria, BC V9E 2E7, Canada}
\altaffiltext{12}{Department of Astronomy, Universidad de Concepci\'{o}n, Casilla 160-C, Concepci\'{o}n, Chile}
\altaffiltext{13}{Jet Propulsion Laboratory, California Institute of Technology, MS 169-327, 4800 Oak Grove Drive, Pasadena, CA 91109, USA}
\altaffiltext{14}{National Astronomical Observatory of Japan, Mitaka, Tokyo 181-8588, Japan}
\altaffiltext{15}{Institute for the Physics and Mathematics of the Universe, The University of Tokyo, 5-1-5 Kashiwanoha, Kashiwa-shi, Chiba 277-8583, Japan}
\altaffiltext{$\star$}{Visiting Astronomer, Kitt Peak National Observatory, National Optical Astronomy Observatories, which is operated by the Association of Universities for Research in Astronomy, Inc. (AURA) under cooperative agreement with the National Science Foundation.}

\begin{abstract}

We have derived masses and ages for 79 early-type galaxies (ETGs) in different environments at $z \sim 1.3$ in the Lynx supercluster and in the GOODS/CDF-S field  using multiwavelength (0.6-4.5 $\mu$m; \textit{KPNO}, \textit{Palomar}, \textit{Keck}, \textit{HST}, \textit{Spitzer}) datasets.  
At this redshift the contribution of the TP-AGB phase is important for ETGs, and the mass and age estimates depend on the choice of the stellar population model used in the spectral energy distribution fits.
We describe in detail the differences among model predictions for a large range of galaxy ages, showing the dependence of these differences on age.
Current  models still yield large uncertainties.
While recent models from Maraston and Charlot \& Bruzual  offer better modeling of the TP-AGB phase with respect to less recent Bruzual \& Charlot models, their predictions do not often match.
The modeling of this TP-AGB phase has a significant impact on the derived parameters for galaxies observed at high-redshift. Some of our results do not depend on the choice of the model:  for all models, the most massive galaxies are the oldest ones, independent of the environment.
When using Maraston and Charlot \& Bruzual models, the mass distribution is similar in the clusters and in the groups, whereas in our field sample there is a deficit of massive ($M \gtrsim 10^{11} M_\odot$) ETGs.
According to those last models, ETGs belonging to the cluster environment host on average older stars with respect to group and field populations.
This difference is less significant than the age difference in galaxies of different masses.\\
\end{abstract}

\keywords{galaxies: clusters: individual (RX J0849+4452, RX J0848+4453)  -- galaxies: elliptical and lenticular -- galaxies: evolution -- galaxies: formation -- galaxies: high-redshift -- galaxies: photometry}

%@@@@@@@@@@@@@@@@@@@@@@@@@@@@@@@@@@@@@@@@@@@@@@@@@@@@@@@@@@@@@@@@@@@@@@@@
% INTRODUCTION
%@@@@@@@@@@@@@@@@@@@@@@@@@@@@@@@@@@@@@@@@@@@@@@@@@@@@@@@@@@@@@@@@@@@@@@@@

\section{Introduction}

Superclusters of galaxies are the largest structures observed in the Universe.
Their dimensions can range between 10 and $\sim$100 Mpc and are composed of two or more galaxy clusters and surrounding groups.
These structures span a large range in galaxy projected number density and permit us to study galaxies seen at the same epoch but in very different environments.
The study of superclusters at high redshifts gives us deep insight into the role of environment in clusters and groups in the very early stages of cluster assembly. 
Deep multiwavelength surveys focusing on the study of superclusters at $z < 1$  \citep[e.g., CL1604 and the ORELSE program; ][]{gal08,lubin09} have shown in detail the variations in star formation rates and galaxy populations as a function of environments. 
It has only been in recent years that superclusters have been discovered at redshift $z>1$ \citep{nakata05,tanaka09}.
In this paper we study the early-type galaxy (ETG) populations in one of those superstructures, the Lynx supercluster. 

Lynx is a high redshift (z $\sim$ 1.26) supercluster, composed of two clusters (RX J0849+4452, hereafter Lynx E and RX J0848+4453, hereafter Lynx W) and surrounding groups \bibpunct[; ]{(}{)}{;}{a}{}{;} \citep[][Mei et al., \textit{in preparation} -- hereafter M11]{nakata05}.
The Lynx W cluster was first identified by \citet{stanford97} as an overdensity in a near-infrared (NIR) imaging survey, and then spectroscopically confirmed at $z_{spec} = 1.273$; it was later detected in a deep Chandra observation \citep{stanford01}. 
The Lynx E cluster was initially found by \citet{rosati99} as a faint extended X-ray source in the ROSAT Deep Cluster Surveys (RCDS) and then spectroscopically confirmed  at $z_{spec} = 1.261$.  The two clusters present different structures.
Lynx E shows a compact galaxy distribution, with a central bright galaxy merger \citep{yamada02,mei06}, while 
Lynx W appears to be at an earlier stage of assembly, with a lack of a clear cD, and galaxies distributed less concentrated.
Their X-ray emission gives luminosities of $L_X^{bol}=(2.8 \pm 0.2) \times 10^{44}$ erg.s$^{-1}$ and $L_X^{bol}=(1.0 \pm 0.7) \times 10^{44}$ erg.s$^{-1}$, for Lynx E and Lynx W respectively \citep{rosati99,stanford01,ettori04}. Their spatial distribution, more compact for Lynx~E and more elongated for Lynx~W, also  indicates that Lynx E is likely more dynamically evolved.
Estimates from \citet{jee06} weak lensing analysis indicate velocity dispersions of  $\sigma = 740^{+113}_{-134}$ km.s$^{-1}$ and $\sigma = 762^{+113}_{-133}$ km.s$^{-1}$ for Lynx E and Lynx W, respectively.
Those values are consistent with the spectroscopic measurements of $\sigma = 720 \pm 140$ km.s$^{-1}$ for Lynx E \citep{jee06} and $\sigma = 650 \pm 170$ km.s$^{-1}$ for Lynx W  \citep{stanford01} \citep[see also][Table 1]{mei09}.

Around the two clusters, \citet{nakata05} identified seven group candidates, using photometric redshifts derived from optical imaging with the \textit{Subaru}/SuprimeCam  \citep{miyazaki02}, indicating the presence of a supercluster extending over an angular distance of $\sim 20$\arcmin, which corresponds to a luminosity distance of $\sim$10 Mpc at the clusters' redshift.
Optical spectroscopy from the \textit{Subaru}, \textit{Keck} and \textit{Gemini} telescopes have confirmed three of those group candidates as being part of the supercluster, with redshifts of $z_{spec} = 1.266 \pm 0.005$ (Group~1), $z_{spec} = 1.262 \pm 0.005$ (Group~2), $z_{spec} = 1.264 \pm 0.003$ (Group~3) (M11). \\

Superclusters host large populations of ETGs that are mostly found in high density regions.
Even if ETGs are relatively simple galaxies, large bulges dominated by dispersion velocities, their formation and evolution is still not well understood. 
In the local Universe they define a very tight red sequence in the color-magnitude space \citep[e.g.,][]{bernardi03d,baldry04}, that begins to dissipate only at the highest redshifts of known clusters  \citep[e.g.,][]{kodama07}.
This suggests that galaxy star formation was quenched in the past and bluer galaxies migrate to the red sequence. 
Multi-wavelength studies of ETGs on the red sequence in different environments give us constrains on the galaxy ages and their star formation history \citep[M11; ][]{rettura10}.

In this paper, we present a deep, panoramic multi-wavelength survey of the Lynx supercluster, ranging from the rest-frame ultraviolet to the infrared.
We will quantify the environmental dependences of the ETG ages and masses, by comparing cluster and group galaxies to a field sample at the same redshift selected from the GOODS (Great Observatories Origins Deep Survey, \citealt{giavalisco04}) observations of the Chandra Deep Field South (CDF-S) \bibpunct[; ]{(}{)}{;}{a}{}{;} \citep[][Dickinson et al., \textit{in preparation}]{giavalisco04,nonino09,retzlaff10} \bibpunct{(}{)}{;}{a}{}{,}.

In $\S$2, we briefly present the observations, the data reduction and the selection of the galaxy sample on which our study relies.
We describe the photometry  in $\S$3 and the spectral energy distribution (SED) fitting method we use to derive stellar masses and ages in $\S$4.
We study in $\S$5 the systematics in the SED fitting, especially the influence of the chosen model.
We then present our results and discuss them in $\S$6 and  $\S$7.
	
In this paper, we adopt a standard cosmology with  $H_0 = 70$ km.s$^{-1}$.Mpc$^{-1}$,  $\Omega_m = 0.30$ and $\Omega_\Lambda = 0.70$.
Unless otherwise specified, all magnitudes are in the AB system and have been corrected for Galactic extinction using the maps of \citet{schlegel98}.

%@@@@@@@@@@@@@@@@@@@@@@@@@@@@@@@@@@@@@@@@@@@@@@@@@@@@@@@@@@@@@@@@@@@@@@@@
% OBSERVATIONS
%@@@@@@@@@@@@@@@@@@@@@@@@@@@@@@@@@@@@@@@@@@@@@@@@@@@@@@@@@@@@@@@@@@@@@@@@

\section{Observations \label{sec:observations}}

	We have obtained images of the Lynx supercluster from the optical to the far-infrared (0.6-4.5 $\mu$m), in seven bandpasses : $R$, \textit{HST}/ACS F775W and F850LP -- hereafter $i_{775}$ and $z_{850}$, $J$, $K_s$, \textit{Spitzer}/IRAC ch1 and ch2 -- hereafter $[$3.6$\mu$m$] $  and $[$4.5$\mu$m$] $.
Information about the data are summarized in \textsc{Tab.}\ref{tab:data_info_clusters} and \textsc{Tab.}\ref{tab:data_info_groups}.

%------------------------------------------------------------------------------------
% OBSERVATIONS: Data reduction
%------------------------------------------------------------------------------------

% HST
	\textit{HST} Advanced Camera for Surveys (ACS) observations of the Lynx superclusters have been carried out as part of the ACS Intermediate Redshift Cluster Survey \citep[Guaranteed Time Observation, or GTO, program 9919; PI: H. Ford;][]{ford04,postman05} for the two clusters and from an \textit{HST}/GO program 10574 (PI: S. Mei) for the groups. 
The \textit{HST} $i_{775}$ and $z_{850}$ cluster (group) imaging was carried out with the ACS Wide Field Camera in March and April 2004 (December 2005 to February 2006)  for a total exposure time of 7290 s and 12220 s (6300 s and 10500 s), respectively.
The ACS WFC resolution (pixel scale) is 0.05\arcsec~ and its field of view is $210~\arcsec \times 240~\arcsec$.
The images were processed with the APSIS pipeline  \citep{blakeslee03a}, with a \textit{Lanczos3}  interpolation kernel.
We adopted AB photometric zeropoints of 25.678 and 24.867 mag respectively in $i_{775}$ and $z_{850}$  from the \textit{HST}/ACS website\footnote{http://www.stsci.edu/hst/acs/}.\\

% KECK & PALOMAR
	The $R$ band images come from two different telescopes.
The clusters have been observed with the \textit{Keck} telescope, and a wider area including Group 1 and Group 2 has been observed with the \textit{Palomar} telescope.
Group 3 has not been covered by our $R$ band imaging.
	
	The \textit{Palomar} $R$ band imaging  (P.I.: D. Stern) was obtained in November 1999 with the COSMIC instrument \citep{kells98} in 18 exposures of 900 s each for a total exposure time of 16200 s.
COSMIC has a resolution of 0.2468 \arcsec/pixel and a field of view of $9.7 \arcmin \times 9.7 \arcmin$.
The images were reduced using standard procedures: the images were bias corrected, trimmed and flat fielded using dome flats and then a super sky flat. 

	The \textit{Keck} $R$ band imaging (P.I.: G. Illingworth) was obtained on November 2003 on a night with photometric conditions on the \textit{Keck I} telescope.
20 exposures of 300-500 s (average 437.5 s) were taken using the red camera on the LRIS Spectrograph \citep{oke95}, for a total exposure time of 8750 s.
LRIS then had a resolution of 0.213\arcsec/pixel and a field of view of $6\arcmin \times 7.8\arcmin$.
The images were reduced using standard techniques.  Each individual image was astrometrically calibrated to the SDSS \citep{york00}.
The final images were then combined using the drizzle software \citep{fruchter02} to a final image scale of 0.1\arcsec/pixel.\\

% KPNO
	The near-infrared $J$ and $K_s$ band imaging (P.I.: A. Gonzalez) was obtained in December 2003 at the \textit{KPNO} 2.1~m telescope with the FLAMINGOS instrument \citep{elston98}.
FLAMINGOS has a resolution of 0.606\arcsec/pixel and a field of view of  $20\arcmin \times 20\arcmin$ on the 2.1~m. 
The imaging was reduced using standard IR imaging reduction techniques, using the DIMSUM package of IRAF scripts. 
The final $J$ (resp. $K_s$) band stacked image was made from about 200 frames of 120 s each (resp. 890 frames of 30 s each) and has a total exposure time of about 24.0 ks (resp. 26.7  ks).\\

% IRAC
	The \textit{Spitzer}/IRAC \citep{fazio98} $[$3.6$\mu$m$] $ and $[$4.5$\mu$m$] $ band imaging of the clusters (resp. groups) was obtained in April 2004 (resp. November 2005 and May 2006) in 30 exposures of 200 s each (resp. 12 exposures of 100 s each) for a total exposure time of 6000 s (resp. 1200 s) (PI: S.A. Stanford).
The data were reduced using standard \textit{Spitzer} procedures. 
The BCD frames were first corrected for muxbleed and pulldown using custom IDL scripts (now available from the Spitzer Science Center), and then processed with MOPEX to produce co-added mosaics.

\begin{deluxetable*}{l c c c c c c c c}
%	\tabletypesize{}
%	\tablewidth{}
%	\tablenum{}
%	\tablecolumns{}
	\tablecaption{Lynx cluster images \label{tab:data_info_clusters}}
	\tablehead{
	\colhead{Image}	& \colhead{Observation}	& \colhead{Telescope}	& \colhead{Instrument} 	& \colhead{Exposure}	& \colhead{FWHM}	& \colhead{Completeness}\tablenotemark{a}\\
	\colhead{} 		& \colhead{date} 		& \colhead{} 			& \colhead{}			& \colhead{time (ks)}		& \colhead{(arcsec)}	& \colhead{(mag)}
		}
	\startdata
	$R$				& Nov. 2003			& \textit{Keck}			&	LRIS	 	   	   	& 8.75	 			& $\sim$ 0.7 		& 26.2		\\
	$i_{775}$		 	& Mar.-Apr. 2004		& \textit{HST}			&	ACS			 	& 7.29 				& $\sim$ 0.1	 	& 26.3		\\
	$z_{850}$		 	& Mar.-Apr. 2004		& \textit{HST}			&	ACS			 	& 12.22 				& $\sim$ 0.1	 	& 25.9		\\
	$J$	 			& Dec. 2003 	 		& \textit{KPNO}			&	FLAMINGOS 	   	& 24	 				& $\sim$ 1.6 		& 21.2		\\
	$K_s	$ 			& Dec. 2003			& \textit{KPNO}			&	FLAMINGOS 	   	& 26.7 				& $\sim$ 1.4 		& 21.1		\\
	$[$3.6$\mu$m$] $	& Apr. 2004 			& \textit{Spitzer}		&	IRAC			& 6	 				& $\sim$ 1.6 		& 23.2		\\
	$[$4.5$\mu$m$] $	& Apr. 2004 			& \textit{Spitzer}		&	IRAC			& 6	 				& $\sim$ 1.6 		& 22.9		\\
	\enddata
	\tablenotetext{a}{50\% point-source completeness for detection at 5$\sigma$}
\end{deluxetable*}

\begin{deluxetable*}{l c c c c c c c c}
%	\tabletypesize{}
%	\tablewidth{}
%	\tablenum{}
%	\tablecolumns{}
	\tablecaption{Lynx group images \label{tab:data_info_groups}}
	\tablehead{
	\colhead{Image}	& \colhead{Observation}	& \colhead{Telescope}	& \colhead{Instrument} 	& \colhead{Exposure}	& \colhead{FWHM}	& \colhead{Completeness}\tablenotemark{a}\\
	\colhead{} 		& \colhead{date} 		& \colhead{} 			& \colhead{}			& \colhead{time (ks)}		& \colhead{(arcsec)}	& \colhead{(mag)}
		}
	\startdata
	$R$				& Nov. 1999			& \textit{Palomar}		&	COSMIC 	   	   	& 16.2	 			& $\sim$ 1.5		& 24.4	\\
	$i_{775}$			& Dec. 2005 - Feb. 2006	& \textit{HST}			&	ACS		 		& 6.3 				& $\sim$ 0.1 		& 26.2	\\
	$z_{850}$			&Dec. 2005 - Feb. 2006	& \textit{HST}			&	ACS		 	  	& 10.5 				& $\sim$ 0.1 		& 25.7	\\
	$J$	 			& Dec. 2003 	 		& \textit{KPNO}			&	FLAMINGOS 	   	& 24 					& $\sim$ 1.6		& 21.2	\\
	$K_s	$ 			& Dec. 2003			& \textit{KPNO}			&	FLAMINGOS 	   	& 26.7 				& $\sim$ 1.4		& 21.1	\\
	$[$3.6$\mu$m$] $	& May - Nov. 2005 		& \textit{Spitzer}		&	IRAC			& 1.2					& $\sim$ 1.6		& 22.2	\\
	$[$4.5$\mu$m$] $ 	& May - Nov. 2005 		& \textit{Spitzer}		&	IRAC			& 1.2					& $\sim$ 1.6		& 21.5	\\
	\enddata
	\tablenotetext{a}{50\% point-source completeness for detection at 5$\sigma$}
\end{deluxetable*}

%------------------------------------------------------------------------------------
% OBSERVATIONS: Sample selection
%------------------------------------------------------------------------------------

\subsection{Sample selection \label{sec:sample_sel}}

Our Lynx sample is an ETG subsample of the cluster and group homogeneous sample described in M11 and \citet{mei06,mei09}.
ETGs have been visually classified in the $B$ rest-frame ($z_{850}$) from \citet{postman05} for the cluster sample and from M11 for the group sample using the same criteria, up to $z_{850}=24$~mag, the limit of reliable visual morphological classification quantified by simulations in \citet{postman05}.
At this magnitude our ACS sample is complete \citep[e.g.,][]{giavalisco04}.
We apply a selection in photometric redshift ($0.92 < z_{phot} <1.36$) and magnitude ($21 \le z_{850} \le 24$~mag).
The $z_{phot}$ are estimated with \textit{Le Phare} \footnote{http://www.oamp.fr/people/arnouts/LE$\_$PHARE.html} \citep{arnouts02,ilbert06} and the selection criteria is calibrated on spectroscopic confirmed members (see M11 for details).
The magnitude cut at $z_{850} = 21$ ensures that no star is included in our sample and the cut at $z_{850} = 24$ secures a reliable morphological classification.
Galaxies belonging to the clusters and groups are then identified by a Friend-Of-Friend algorithm \citetext{FOF, \citealp{geller83}; see also \citealp{postman05}} with a linking scale corresponding to a local distance of $0.54$~Mpc, normalized to $z = 1.26$ and to our magnitude range as in \citet{postman05}.
We use the X-ray emission center for the clusters and the overdensity centers defined by \citet{nakata05} for the groups.
Spectroscopically confirmed outliers were excluded from the sample.

Our CDF-S sample has been selected following similar criteria.
We use as a starting catalog the public GOODS-MUSIC v2 sample \citep{santini09}, which is complete at $z_{850} = 24$ (90\% complete at $z_{850} = 26$) and contains about $\sim$15,000 objects.
Photometric redshifts are available for all objects and spectroscopic redshifts, collected from public surveys, are available for about 2,900 of those objects.
We apply to this CDF-S sample the same magnitude cut in $z_{850}$ band and we select objects with secure $z_{spec}$ (quality flag=0,1) within $1.1 \le z_{spec} \le 1.4$.
By comparing our selection with the number of objects with $21 \le z_{850} \le 24$ and $1.1 \le z_{phot} \le 1.4$, we estimate that our CDFS sample is more than 70\% complete.
ETGs were identified by visual morphological classification in  the $z_{850}$ bandpass, consistent with the Lynx classification.
We also verify that these ETGs are field ETGs, i.e. that they do not belong to already identified structures in the CDF-S.
According to \citet{salimbeni09a}, there are twelve identified structures within the redshift range 0.4-2.5, four of which lie at $z \sim 1.1$.
We exclude from our CDF-S sample one ETG, which most likely belongs to one of those structures.
We check the consistency of our morphology classification with that of \citet{bundy05}: the two classifications agree on all galaxies that are in common ($z_{850} < 22.5$, 6 galaxies).
We thus obtain 27 ETGs in the CDF-S with $\langle z_{spec} \rangle = 1.241 \pm 0.082$.

The Lynx cluster, group and CDFS-S field samples have similar spectral coverage and are complete at $z_{850} = 24$ mag, thus providing an homogeneous and consistent sample.
A possible bias that might affect the CDF-S sample would be a lack of low-mass/passive ETGs that are not included in the spectroscopic sample because of their faint absorption lines.
We will discuss this in our result section.

Our final sample consists of 79 ETGs comprising 31 in the Lynx clusters, 21 in the Lynx groups and 27 in the CDF-S.
Our CDF-S sample and about half of our Lynx sample have spectroscopic redshifts.
We remark that known AGNs have been removed from the sample.
In a companion paper, (Rettura et al., \textit{in preparation} -- hereafter R11), we study in detail the star formation histories of the subsample of 13 massive ($M > 5 \cdot 10^{10} M_\odot$) spectroscopically confirmed ETGs in the Lynx clusters.

%@@@@@@@@@@@@@@@@@@@@@@@@@@@@@@@@@@@@@@@@@@@@@@@@@@@@@@@@@@@@@@@@@@@@@@@@
% PHOTOMETRY
%@@@@@@@@@@@@@@@@@@@@@@@@@@@@@@@@@@@@@@@@@@@@@@@@@@@@@@@@@@@@@@@@@@@@@@@@

\section{Photometry \label{sec:photometry}}
	
In this paper we will use AB magnitudes in all bandpasses.
\textsc{Tab.}\ref{tab:vega2ab} gives the magnitude conversion between Vega and AB system.
For ACS and IRAC, those values come from the websites\footnote{http://www.stsci.edu/hst/acs/analysis/zeropoints}$^,$\footnote{http://web.ipac.caltech.edu/staff/gillian/cal.html}.
For ground-based telescopes, those values have been estimated using the Vega spectrum given by \citet{kurucz93}\footnote{ftp://ftp.stsci.edu/cdbs/grid/k93models/standards/vega\_k93.fits}.
\begin{deluxetable}{l c}
%	\tabletypesize{}
	\tablewidth{\linewidth}
%	\tablenum{}
%	\tablecolumns{}
	\tablecaption{Magnitude system conversion ($\Delta = \;$AB - Vega) \label{tab:vega2ab}}
	\tablehead{
		\colhead{Image}	&	$\Delta = \;$AB - Vega	\\
		\colhead{}			&	\colhead{(mag)}
	}
	\startdata
$R$	\textit{(Keck)}	&	0.19	\\
$R$	\textit{(Palomar)}&	0.22	\\
$i_{775}$			&	0.39	\\
$z_{850}$			&	0.52	\\
$J$				&	0.92	\\
$K_s$			&	1.89	\\
$[$3.6$\mu$m$] $	&	2.79	\\
$[$4.5$\mu$m$] $	&	3.26	\\
	\enddata
%	\tablecomments{}
\end{deluxetable}

	While SExtractor provides excellent source detection and generally good photometry, it has been found \citep[e.g.,][]{giavalisco04, blakeslee06, haussler07, mei09} that there are some systematics in SExtractor's photometry due to sky overestimation.
We find similar systematics for fixed aperture photometry : when setting the inner radius of the annulus used for the sky estimation (with the \texttt{BACKPHOTO\_TYPE} keyword set to \texttt{LOCAL}), SExtractor uses the extension of the source defined by SExtractor segmentation map and multiplies it by 1.5.
In the \textit{HST}/ACS images, because of the small PSF FWHM, this inner radius is in general smaller than the aperture radius and this often leads to overestimating the sky, because there is still non negligible light from the galaxy in the sky annulus.

	In order to take into account the large range of PSF FWHMs spanned by our dataset, we perform matched aperture photometry (see below), with an aperture radius of 1.5\arcsec~ and an aperture correction out to 7\arcsec~ radius \citep[see also][]{rettura06}.
The aperture radius of 1.5\arcsec, close to the maximum extension of our PSF FWHMs (IRAC), is a compromise between maximizing the flux of the source and minimizing the contamination by other sources and sky.
The aperture correction radius, 7\arcsec, is also a compromise between those two opposing goals.\\

	To obtain accurate photometry, we first build a mask for neighboring objects.
To this end, we use masks obtained with the software SExtractor (we beforehand corrected the cases when SExtractor attributes multiple detections to the studied galaxy). 
We then perform aperture photometry:  we estimate and subtract the sky, count the flux within the aperture radius and, eventually, apply the aperture correction in order to take into account the flux outside the aperture radius.\\

%------------------------------------------------------------------------------------
% PHOTOMETRY: Sky estimation
%------------------------------------------------------------------------------------

\subsection{Sky estimation}	
	Sky estimation is a key step in photometry, because it can significantly change the magnitude (in some cases, magnitudes derived with SExtractor's sky estimation can differ up to $\sim$0.2 from our magnitudes).
Due to the variety of our dataset, we set up our own sky estimator. 
We first mask the objects with ellipses, using SExtractor's structural parameters linearly increased by a factor of 1.5.
We make a first estimate of the sky by taking the median value, called $sky_0$, in a 3\arcsec~ thick circular annulus.  The inner radius is set as the maximum between 1.5\arcsec~ (the aperture radius) and 1.5 times the extension of the source.
We check the galaxy growth curve thus obtained: if it decreases, meaning that we overestimate the sky (usually because of residual light from masked close bright objects), we reduce the sky value by 10\% of $sky_0$; we repeat this step until the growth curve flattens. 
This method allows us to detect and  correct the cases when the sky has been overestimated, without any assumption about the shape of the growth curve.
We do not correct the cases where the sky has been underestimated because their detection requires assumptions about the growth curve and they are rare.\\

%------------------------------------------------------------------------------------
% PHOTOMETRY: Aperture correction
%------------------------------------------------------------------------------------

\subsection{Aperture correction \label{sec:ap_corr}}	
	We measure the flux within a 1.5\arcsec aperture radius.
We then make an aperture correction out to 7\arcsec, i.e. we assume that all the flux of the galaxy is enclosed in a 7\arcsec~ radius circle.
Aperture correction is usually done using the PSF growth curve.
Since the difference between the observed ETG growth curve and the PSF growth curve can be significant for the \textit{HST}/ACS images, we estimate this difference using simulations.
In order to build a growth curve, we simulate 1,000 galaxies. 
The simulated galaxies have the following characteristics representative of our sample: a \citet{sersic68} profile  with an index $n_{ser}$, an axis ratio $b/a$ and an effective radius $r_e$ to which we attribute a normal distribution ($n_{ser}=4 \pm 2$, $b/a = 0.65 \pm 0.2$ and $r_e = 0.25\arcsec \pm 0.15\arcsec$) and a random position angle $p.a.$.
We then convolve the simulations with the PSF (created from selected stars, normalized at 7\arcsec~ radius) and add a Poissonian distribution.
	
\begin{figure}
	\includegraphics[width=\linewidth]{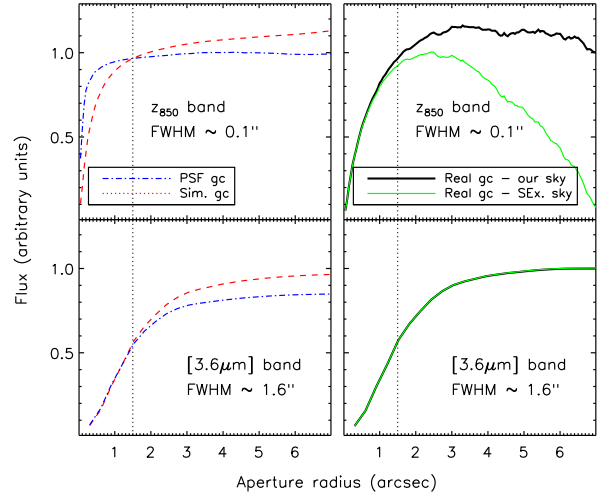}
	\caption{
\textit{Left panels}: comparison between the PSF (blue dash-dotted line) and simulated galaxy (red dashed line) growth curves;
if the PSF FWHM is small compared to the galaxy size (\textit{HST}/ACS), the PSF and the simulated galaxy growth curves differ significantly at small radii, though it is not the case if the FWHM is greater than the galaxy size (\textit{Spitzer}/IRAC).
\textit{Right panels}: example of a Lynx galaxy growth curve using our sky estimate (thick black solid line) and using SExtractor's sky estimate (light green solid line); 
we observe that SExtractor can clearly overestimate the sky in the \textit{HST}/ACS bandpasses.\\}
	\label{fig:gal_gc}
\end{figure}

	In \textsc{Fig.} \ref{fig:gal_gc}, we show (left panels) the difference between the PSF and simulated galaxy growth curves for the two extreme (smallest and biggest) PSF FWHM in our dataset (ACS/\textit{HST} z$_{850}$ (top panels) and \textit{Spitzer}/IRAC $[$3.6$\mu$m$] $ (bottom panels)).
We can see on the one hand that, for the \textit{HST}/ACS band (FWHM $\sim$ 0.1\arcsec), the simulated galaxy growth curve is much closer, at small radii, to the real galaxy growth curve than the PSF one.
On the other hand, for the \textit{Spitzer}/IRAC band (FWHM $\sim$ 1.6\arcsec), the simulated galaxy growth curve is very similar to the PSF growth curve.
However, the difference in total flux between the simulated galaxy and the PSF growth curves is similar in the two bandpasses.

	On the right panels of \textsc{Fig.} \ref{fig:gal_gc}, we show a Lynx galaxy growth curve calculated using our sky estimate (thick black solid line) and the SExtractor's sky estimate (light green solid line): this illustrates how SExtractor can overestimate the sky in the \textit{HST}/ACS images.
Pixel-to-pixel sky variations in the \textit{HST}/ACS sky value are high, which make the growth curve less regular.

% FIGURE 1	

	Aperture corrections, using PSF and simulated ETG growth curves, are displayed in \textsc{Tab.}\ref{tab:apcor}.
Interestingly, we observe two effects.
First, the difference between the two aperture corrections is relatively independent of the image PSF FWHM: it is roughly a difference of 0.1-0.15 magnitude.
Second, the uncertainty on the  aperture correction using the simulated galaxies is equal to 0.08 mag independent of the band.
In order to understand these two points, we study the dependence of the aperture correction on $r_e$, $n_{ser}$ and $b/a$.
We observe that the aperture correction is strongly correlated with $r_e$, weakly correlated with $n_{ser}$ and independent of $b/a$.
In order to alleviate the dependence on $n_{ser}$, we repeat our simulations with the Sersic index fixed at $n_{ser} = 4$ \citet{de-vaucouleurs48} profile.
We obtain values similar to the ones in the right column of \textsc{Tab.}\ref{tab:apcor}, but with a slightly lower dispersion (0.06 instead of 0.08).

\begin{figure*}
 	\begin{tabular}{ccc}
	   \includegraphics[width=0.5\linewidth]{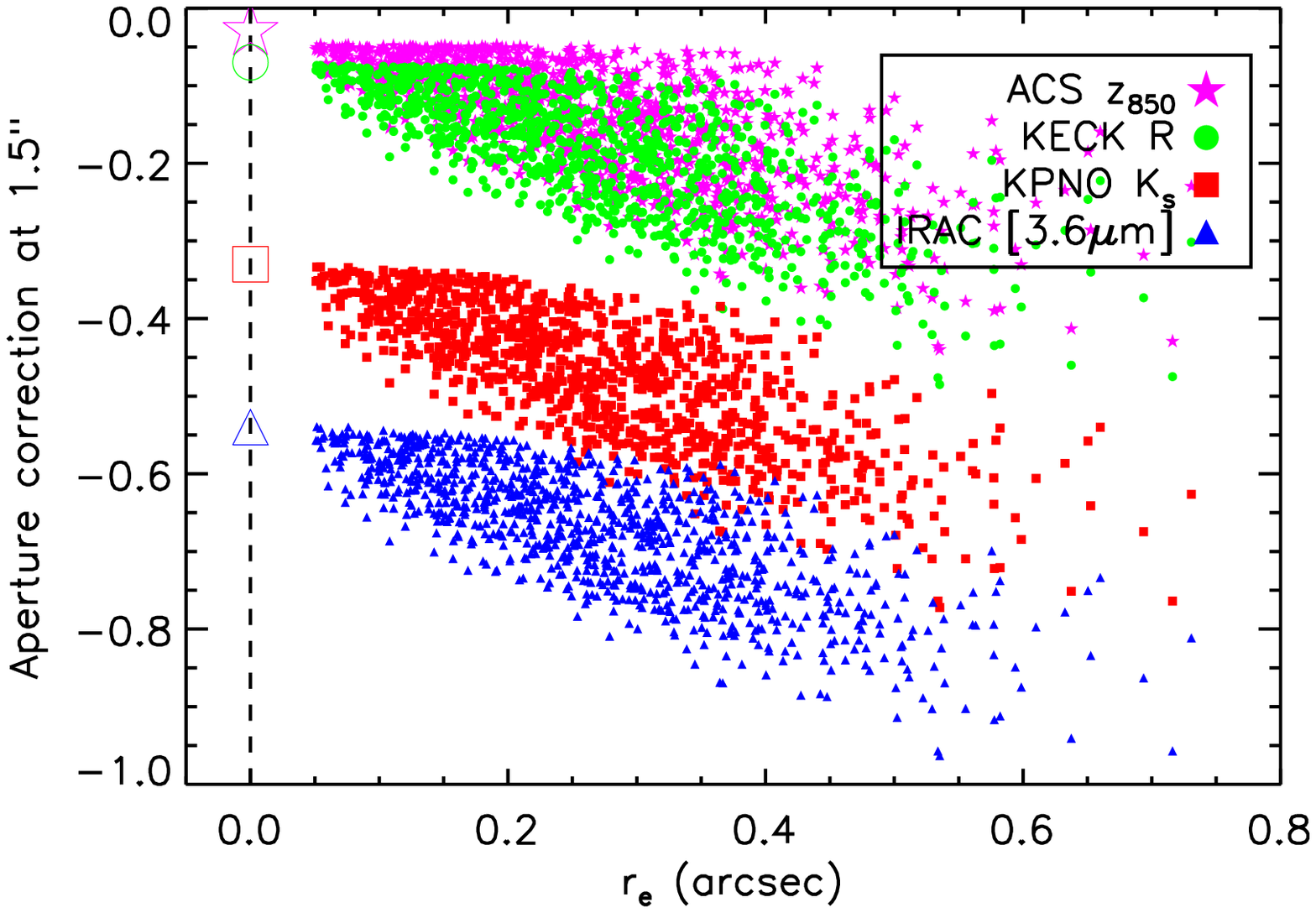} & 
	   \includegraphics[width=0.5\linewidth]{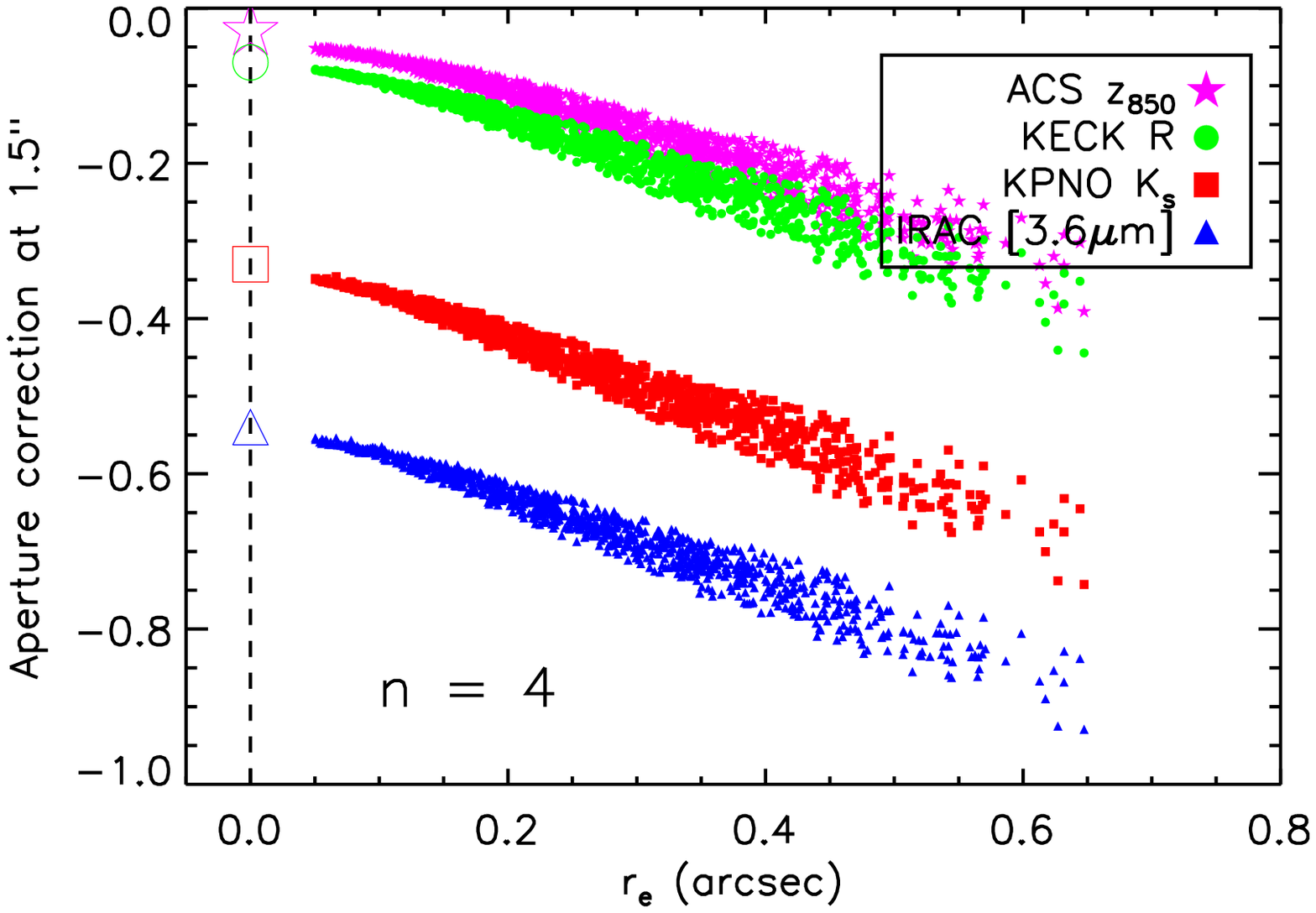} \\
	\end{tabular}
	\caption{Aperture correction as a function of the effective radius $r_e$ for simulated ETGs.
The simulated ETGs have a Sersic index following a normal distribution ($n_{ser} = 4 \pm 2$) on the left panel and a fixed Sersic index ($n_{ser} = 4$) on the right panel.
For four bandpasses (\textit{HST}/ACS $z_{850}$: magenta stars, \textit{Keck} $R$: green disks, \textit{KPNO} $K_s$: red squares, IRAC $[3.6\mu$m$]$: blue triangles) and for each simulated ETG, we represent the aperture correction -- i.e. the magnitude correction corresponding to the flux outward of 1.5\arcsec-- as a function of $r_e$.
At $r_e = 0$, we show (large empty symbols) the aperture correction corresponding to the PSF growth curve (left column of \textsc{Tab.}\ref{tab:apcor}).\\}
	\label{fig:apcor}
\end{figure*}

	We show in \textsc{Fig.}\ref{fig:apcor} how the aperture correction for simulated galaxies -- convolved with the PSF -- depends on the effective radius $r_e$.
The left panel shows the values for four representative bandpasses of our dataset (\textit{HST}/ACS $z_{850}$: magenta stars, \textit{Keck} $R$: green dots; \textit{KPNO} $K_s$: red squares; \textit{Spitzer}/IRAC $[$3.6$\mu$m$]$: blue triangles).
The right panel shows the same, but the simulated galaxies have a fixed Sersic index ($n_{ser} = 4$).
The large empty symbols at $r_e = 0$ represent the aperture correction derived from the PSF growth curve.

% FIGURE 2

This figure illustrates that the dispersion on the aperture correction mainly results from the dispersion on $r_e$ for the simulated galaxies and explains why this dispersion is independent on the considered band.
Moreover, it allows us to understand why the difference between the two aperture correction methods (PSF and simulated galaxies) is relatively independent of the considered band: the properties of each image are included in the PSF, hence the variation with $r_e$ of the amount of galaxy light outward of 1.5\arcsec~ is independent of the PSF.

	In this work, we choose to use aperture corrections derived from the PSF growth curve (left column of \textsc{Tab.}\ref{tab:apcor}) in order to facilitate the comparison with other studies. 
The photometric and fitted parameters catalogs in Appendix  are built with the PSF growth curve aperture correction.
In $\S$5, we analyze the impact of the choice of aperture correction on the fitted parameters (stellar mass, age).
The aperture corrections derived from the simulated galaxies growth curve are given in \textsc{Tab.}\ref{tab:apcor}.

\begin{deluxetable}{l c c }
%	\tabletypesize{}
	\tablewidth{\linewidth}
%	\tablenum{}
%	\tablecolumns{}
	\tablecaption{PSF and simulated galaxy aperture corrections \label{tab:apcor}}
	\tablehead{
		\colhead{Image} & \colhead{PSF}	 & \colhead{Simulated galaxies}\\
		\colhead{}		&	\colhead{(mag)}	&	\colhead{(mag)}
		}
	\startdata
	$R$ (\textit{Keck})	& -0.07 $\pm$ 0.02 	& -0.18 $\pm$ 0.08 \\
	$R$ (\textit{Palomar})& -0.33  $\pm$ 0.05	& -0.46 $\pm$ 0.08  \\
	$i_{775}$ 		   	& -0.05 $\pm$ 0.01 	& -0.15 $\pm$ 0.08 \\
	$z_{850}$ 	   	& -0.03 $\pm$ 0.01 	& -0.12 $\pm$ 0.08  \\	
	$J$ 				& -0.33 $\pm$ 0.03 	& -0.47 $\pm$ 0.08  \\
	$K_s	$ 			& -0.33 $\pm$ 0.03 	& -0.47 $\pm$ 0.08  \\
	$[$3.6$\mu$m$] $ 	& -0.54 $\pm$ 0.02 	& -0.67 $\pm$ 0.08  \\
	$[$4.5$\mu$m$] $ 	& -0.56 $\pm$ 0.03 	& -0.69 $\pm$ 0.08  \\
	\enddata
	\tablecomments{The aperture radius is 1.5\arcsec~ and growth curves have been normalized at 7\arcsec. Throughout all this article, we use the PSF correction.}
\end{deluxetable}

%------------------------------------------------------------------------------------
% PHOTOMETRY: Magnitude errors
%------------------------------------------------------------------------------------

\subsection{Magnitude errors \label{sec:mag_err}}	

	In order to estimate the error made on the measured magnitudes, we used Monte Carlo simulations.
For each band and each magnitude in the magnitude range of our Lynx galaxies with a step of 0.2 magnitude, we simulate 1,000 galaxies with the characteristics described in $\S$\ref{sec:ap_corr}.
We add simulated galaxies in a real image at a random position with no object within a 1\arcsec~ radius circle and measure their magnitudes with the same method as described above for the observations.
So, for a given band and a given input magnitude, we now have 1,000 values of simulated measured magnitudes.
We then estimate the magnitude uncertainty as their standard deviation calculated with 3$\sigma$-clipping iterations.
To this estimated uncertainty, we add in quadrature a zeropoint uncertainty (0.01 mag for \textit{HST}, 0.03 mag for \textit{KPNO}, 0.04 mag for \textit{Keck} and \textit{Palomar}, 0.05 mag for \textit{Spitzer}).
For the PSF aperture correction, we also add quadratically the uncertainty on the aperture correction (left column of \textsc{Tab}.\ref{tab:apcor}).
If the simulated galaxies growth curve is used, this uncertainty is already included in the magnitude uncertainty derived above.

%@@@@@@@@@@@@@@@@@@@@@@@@@@@@@@@@@@@@@@@@@@@@@@@@@@@@@@@@@@@@@@@@@@@@@@@@
% SED FITTING
%@@@@@@@@@@@@@@@@@@@@@@@@@@@@@@@@@@@@@@@@@@@@@@@@@@@@@@@@@@@@@@@@@@@@@@@@

\section{SED Fitting \label{sec:sedfit}}

	Our catalog permits us to build SED, from which we can derive basic galaxy properties.
We fit the galaxy SED using stellar population synthesis models to obtain stellar masses and stellar population ages.

%------------------------------------------------------------------------------------
% SED FITTING: Models used
%------------------------------------------------------------------------------------

\subsection{Models used}

	We compare the results using composite stellar population synthesis models from \citet{bruzual03} (BC03), \citet{maraston05} (MA05) and Charlot \& Bruzual (CB07), with a  Salpeter IMF \citep{salpeter55}, solar metallicity, no dust and exponentially declining star-formation history $\psi$ (characteristic time SFH $\tau$: $\psi(t) \propto e^{-t/\tau}$).
We thus have a 3D $\lbrace$SFH $\tau$, age, mass$\rbrace$ grid of synthetic spectra to compare with our measurements.
The fitting parameters and their range are listed in \textsc{Tab.}\ref{tab:fit_param}.

	Because of the age-metallicity degeneracy and to reduce the number of free parameters, we keep the metallicity fixed during the fit.
In $\S$\ref{sec:influence_fit_others}, we discuss how results change when changing the IMF of introducing dust.
The choice of a SFH as simple as an exponentially declining SFH is justified by the lack of information on rest-frame UV emitted light (see hereafter), which prevents to constrain more complex SFHs.

	We note that the stellar mass is the mass locked into stars, including stellar remnants (column 7 of \textit{*.4color} files for BC03/CB07 models and "M $\wedge \ast$ total" for MA05 models).
The best-fit age output by the models is the time elapsed since the onset of star formation.
A more meaningful age is the star-formation weighted age, which represents the age of the bulk of the stars.
If we note $\langle t \rangle_{SFW}$ the star-formation weighted age and $t$ the best-fit age output by the models, we have for an exponentially declining star-formation history $\psi$:
\begin{equation}
\langle t \rangle_{SFW} = \frac{\int_0^t (t-t') \cdot \psi(t') \cdot dt'}{\int_0^t \psi(t') \cdot dt'} = \frac{t-\tau+\tau \cdot e^{-\frac{t}{\tau}}}{1-e^{-\frac{t}{\tau}}}
\end{equation}
Throughout this work, we will consider the star-formation weighted age.

\begin{deluxetable}{l l }
%	\tabletypesize{}
	\tablewidth{\linewidth}
%	\tablenum{}
%	\tablecolumns{}
	\tablecaption{Parameters used for fitting \label{tab:fit_param}}
	\tablehead{
		\colhead{Parameter} & \colhead{Settings}
		}
	\startdata
	IMF				& Salpeter - fixed\\
	Metallicity			& Solar - fixed\\
	Dust				& No dust - fixed\\
	SFH $\tau$ (Gyrs)	& [0.1, 0.25, 0.4, 0.5, 0.75, 1, 1.5, 2, 3, 4, 5] \\
	Age (Gyrs)		& [0.1, age of the Universe at the considered redshift]\\
	Mass(M$_{\sun}$)	& 10$^8$ $\le$ M $\le$10$^{12}$\\
	\enddata
\end{deluxetable}

%------------------------------------------------------------------------------------
% SED FITTING: Fitting method, uncertainties estimations
%------------------------------------------------------------------------------------

\subsection{Fitting method, uncertainties estimations \label{sec:sed_fit_err}}

	We derive the  $\lbrace$SFH $\tau$, age, mass$\rbrace$  parameters by choosing the combination which minimizes the $\chi^2$ defined by:
\begin{equation}
\chi^2 = \sum_{i=1}^{n} \left( \frac{m_{i,data}-m_{i,model}}{\sigma_i}\right)^2,
\end{equation}
where the index $i$ denotes the band, $m_{i,*}$ the magnitude in this band and $\sigma_i$ the associated uncertainty.
For estimating the 1$\sigma$ confidence intervals, we use the method suggested by \citet{papovich01}.
For each galaxy, we simulate 250 sets of SEDs by perturbing the original photometry randomly within their uncertainties.
We fit each set with the above method, thus having 250 values of minimum $\chi^2_{min}$.
We take the value $\chi^2_{conf}$ for which we have $\chi^2_{min} < \chi^2_{conf}$ for 68\% of the simulated cases.
The 1$\sigma$ confidence interval for the original fit is the portion of the $\chi^2$ surface where $\chi^2 < \chi^2_{conf}$.

To test how the lack of the measurement in one bandpass affects our final results, we have performed a simple test on the subsample of galaxies detected in all bandpasses.
When performing the SED fitting on a subsample of magnitudes available for each galaxy, we obtain stable results only when at least the ACS and one NIR ($J,K_s$) or IRAC bandpasses are available.
From hereafter, we only consider galaxies that have photometry at least in these bandpasses.
We thus exclude 4 galaxies for the clusters and 1 for the groups: those galaxies are very close to another object, which cannot be deblended in any of the NIR and IRAC bandpasses.

	We do not consider our SFH $\tau$ estimates accurate, because, as mentioned in \citet{rettura10}, rest-frame UV photometry ($\lambda_{rest} < 200$ nm) is needed for precisely determining the SFH $\tau$.
In this article, we will restrict our analysis to the age and stellar mass; star formation histories for cluster ETGs will be studied in our companion paper R11.

\begin{figure*}
 	\begin{tabular}{ccc}
	   \includegraphics[width=0.5\linewidth]{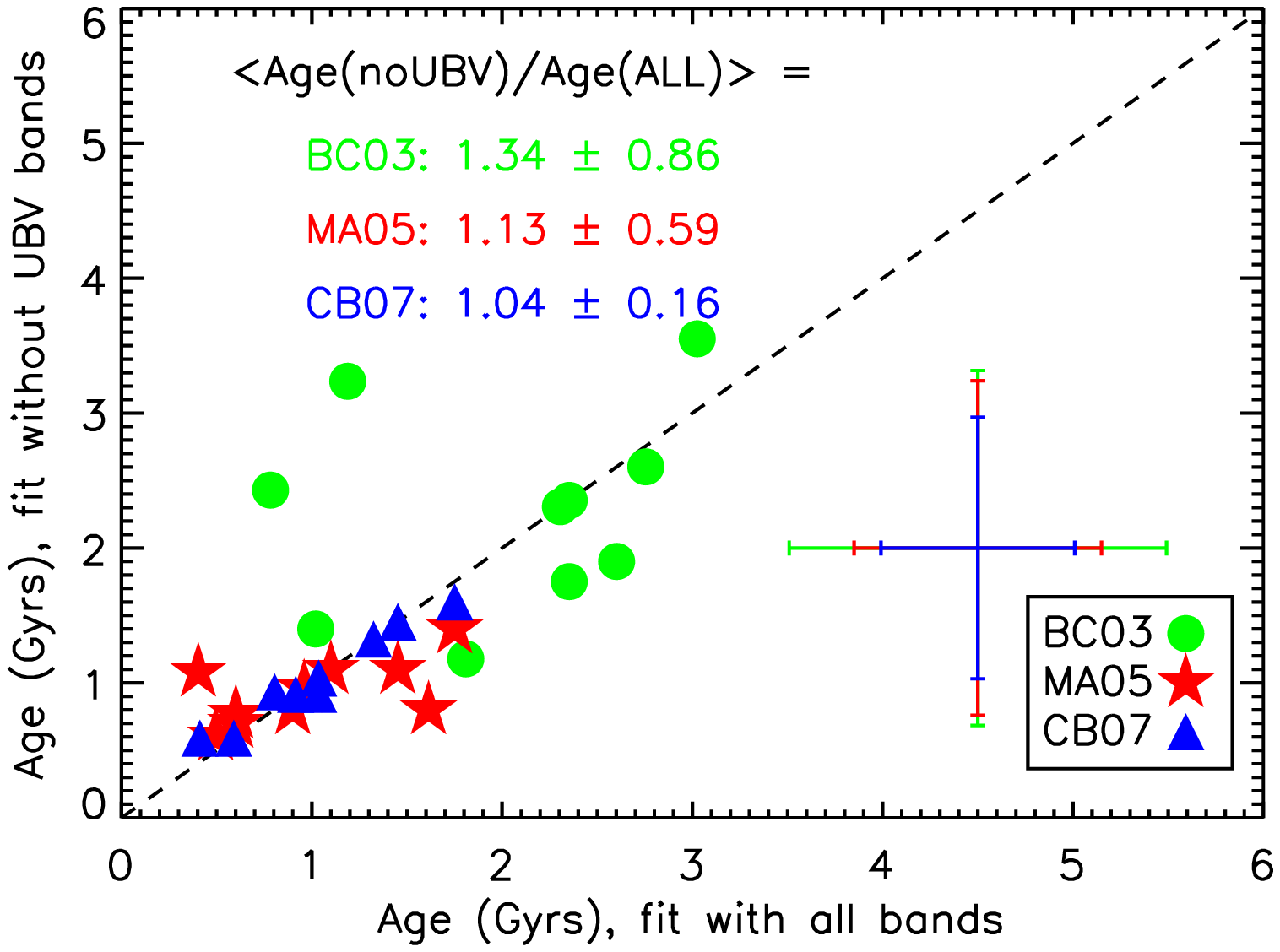} & 
	   \includegraphics[width=0.5\linewidth]{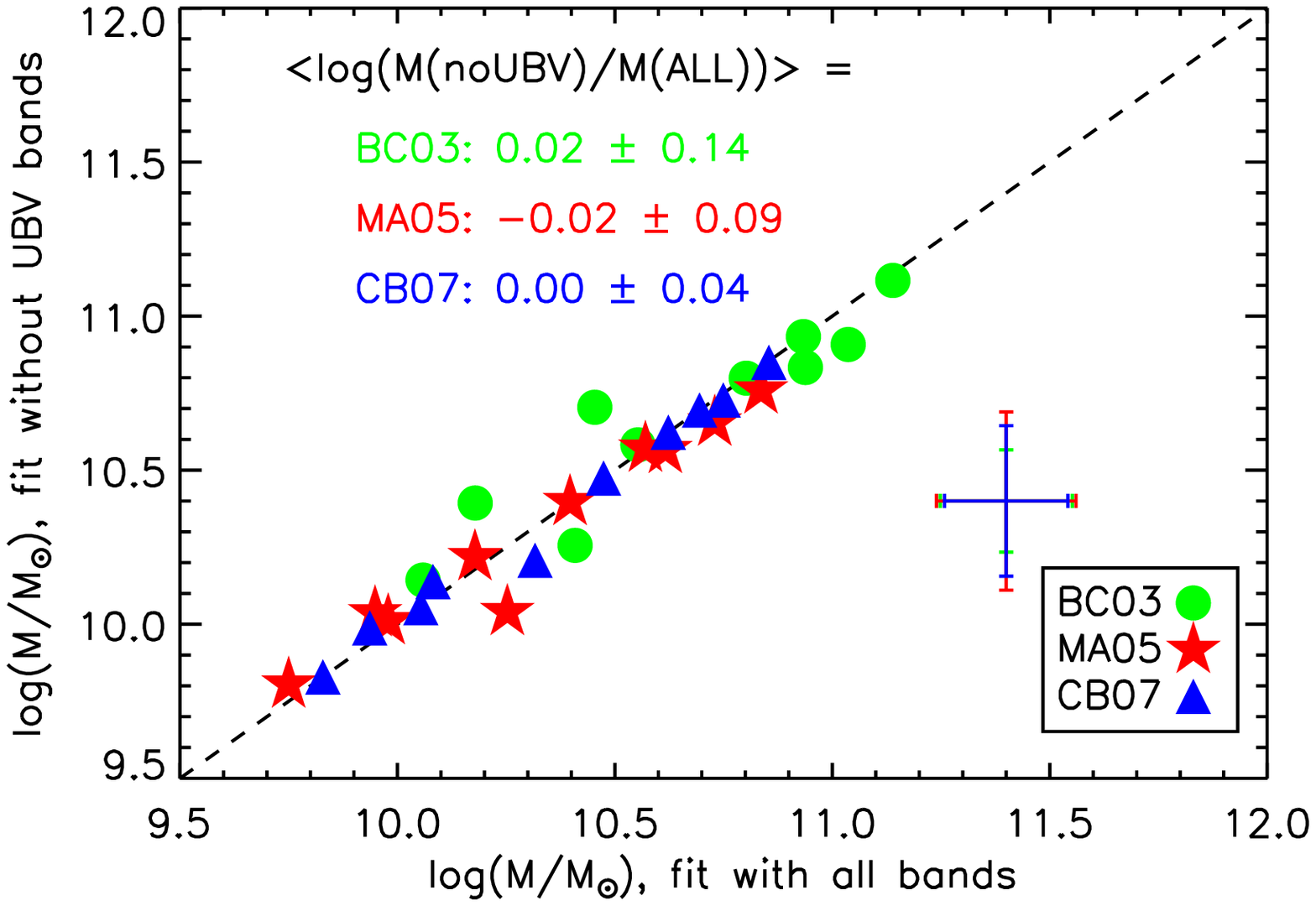} \\
	\end{tabular}
	\caption{Reliability of the estimated ages and stellar masses: for a subset of 10 ETGs from our CDF-S set for which the photometry is available in the 10 bandpasses from the $U$ band to the [$4.5\mu$m] band, we fit the SED once with using all the bandpasses and once without using $UBV$ bandpasses.
We observe that the ages and masses derived without $UBV$ bandpasses are overall in agreement with those derived with $UBV$ bandpasses within the errors.
The error bars show the mean error on the estimated age/mass.\\}
	\label{fig:noUBV}
\end{figure*}

	In order to verify that our range in wavelength is suitable to derive dependable ages, we used a test sample.
We consider the subset of 10 ETGs in our CDF-S set, for which the photometry is available in $UBVRi_{775}z_{850}J_sK_s[3.6\mu$m][$4.5\mu$m].
For this set of galaxies, we fit the SED both using the complete set of bandpasses and only $Ri_{775}z_{850}J_sK_s[3.6\mu$m][$4.5\mu$m].
We then compare the ages and stellar masses we obtain with those two fits.
As it can be seen in \textsc{Fig.}  \ref{fig:noUBV},  the ages derived adding the $UBV$ bandpasses are mostly consistent with those derived using the bandpasses sampled in this paper, though the uncertainty remains large.
We expected this result, because the $i_{775}$ and $z_{850}$ filters bracket the $4000$\AA ~ break and thus constrain the age.
Moreover, as expected, we observe that the masses are not significantly affected by the addition of the $UBV$ bandpasses.
For both ages and masses, adding the $UBV$ bandpasses however reduces  the uncertainties in the derived values.

% FIGURE 3

%@@@@@@@@@@@@@@@@@@@@@@@@@@@@@@@@@@@@@@@@@@@@@@@@@@@@@@@@@@@@@@@@@@@@@@@@
% INFLUENCE OF THE MODELS AND FIT PARAMETERS
%@@@@@@@@@@@@@@@@@@@@@@@@@@@@@@@@@@@@@@@@@@@@@@@@@@@@@@@@@@@@@@@@@@@@@@@@

\section{Study of the systematics in the SED fitting \label{sec:influence_fit}}

%------------------------------------------------------------------------------------
% INFLUENCE OF THE MODELS AND FIT PARAMETERS: Comparison of BC03/MA05/CB07 stellar population synthesis models
%------------------------------------------------------------------------------------

\subsection{Uncertainties due to stellar population synthesis models (BC03/MA05/CB07) \label{sec:influence_fit_models}}
	
The SED fitting results depend on the stellar population synthesis model that was used.
Until recently, the most commonly used models were the BC03.
\citet{maraston06} have shown that the treatment of the TP-AGB phase of stellar evolution is a source of major discrepancy in the determination of the stellar age and mass of high-z galaxies.
The modeling of this phase is challenging:
its rapid evolution makes it difficult to obtain the observations needed to constrain the models and the physics.
Moreover the TP-AGB phase has an important contribution at the redshift of our sample.
BC03 models seem to underestimate the impact of the TP-AGB phase, which explains why the inferred ages/masses are higher than those inferred with MA05.
The underestimation of the TP-AGB phase in the BC03 prescription makes a modeled galaxy less bright and less red in the rest-frame NIR.
To fit the observation, we then need an older and more massive model.
The effect of the TP-AGB phase is dominant during the first Gyrs of the galaxy: taking it into account thus make the galaxy brighter and redder at young ages.
As our ETGs are observed when the Universe is about 5 Gyrs old, this effect is important.
To better take into account this phase, Charlot \& Bruzual implemented in their new models (often referred to as CB07) the results of \citet{marigo08} on the TP-AGB evolutionary phase.
These CB07 models are still in a preliminary version, but the inferred ages and masses are lower than those with BC03 models, similar to MA05 \citep[see for instance][]{bruzual07a}.  
In order to understand how each model changes our inferred masses and ages and to derive conclusions that are independent of the models, we fit our data with BC03, MA05 and CB07 set to the same parameters (see  \textsc{Tab.} \ref{tab:fit_param}).

\begin{figure}
	   \includegraphics[width=0.9\linewidth]{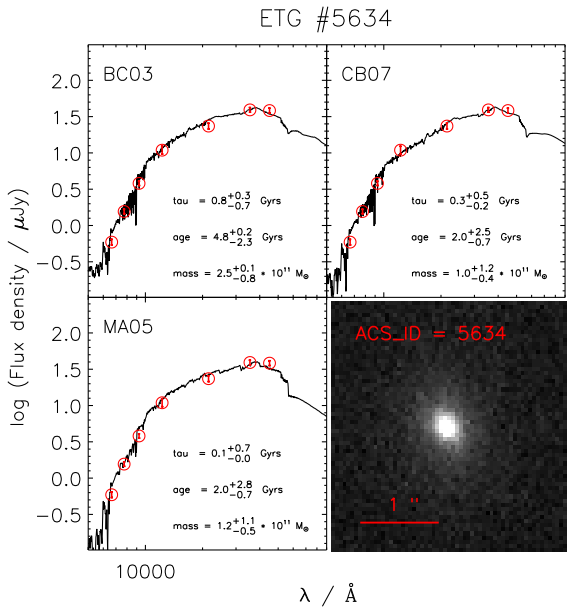}
	\caption{Example of a SED fitting:
in the lower right panel is a stamp (\textit{HST}/ACS $z_{850}$ image) of a Lynx ETG and, 
in each other panel, is plotted in red the observed SED ($Ri_{775}z_{850}JK_s[3.6\mu$m$][4.5\mu$m$]$) of this Lynx ETG.
For each model (BC03/MA05/CB07), the best-fit spectrum and the best-fit parameters are reported.\\}
	\label{fig:sedfit}
\end{figure}

	As an illustration, we show in \textsc{Fig.} \ref{fig:sedfit}, for one ETG of our sample, an \textit{HST}/ACS $z_{850}$ stamp, its SED and the best-fit spectrum and the output parameters derived with each of the three models.
Though the three fits are of similar quality, the derived age and stellar mass are different.

% FIGURE 4

\begin{figure*}
	   \includegraphics[width=\linewidth]{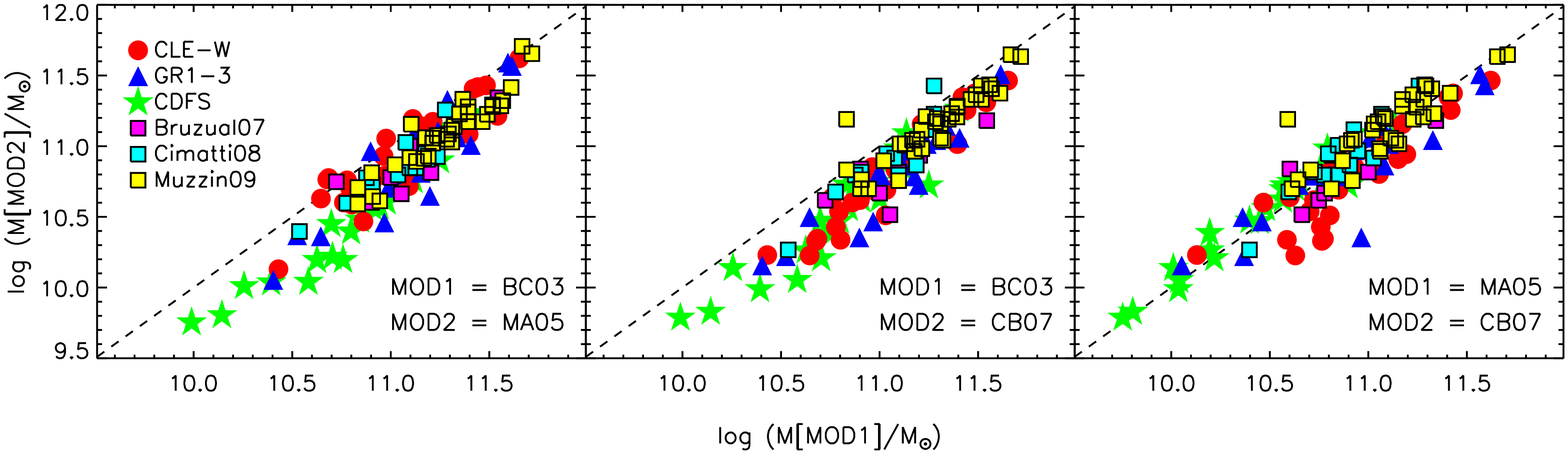}\\
	   \includegraphics[width=\linewidth]{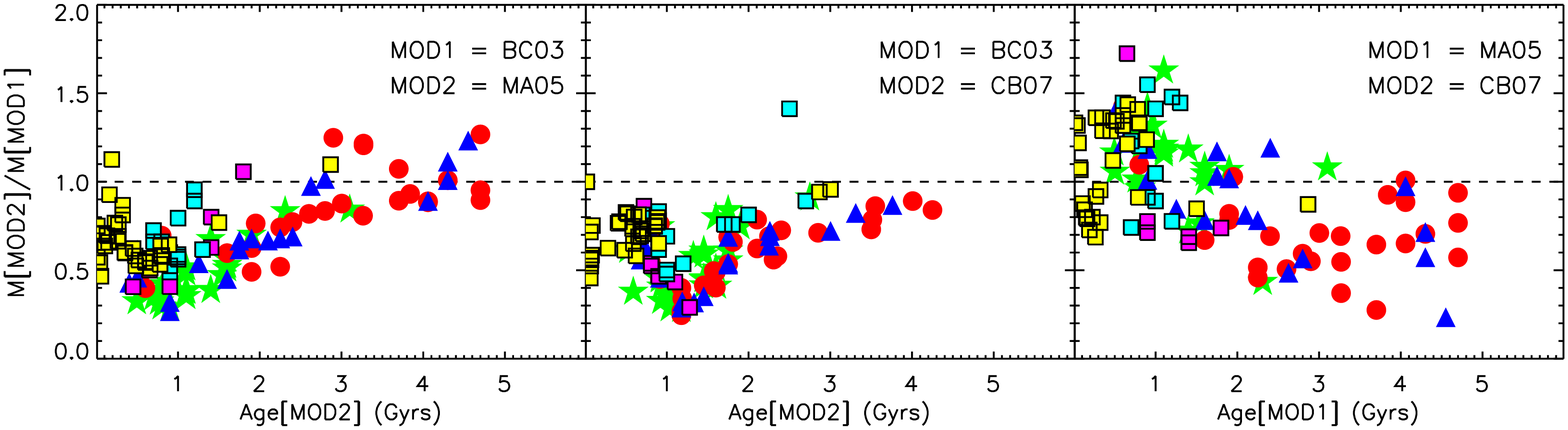}
	\caption{The dependence of the estimated mass on the stellar population models (BC03/MA05/CB07).
\textit{Top}: the mass estimated with one model against the mass estimated with another.
\textit{Bottom}:  the mass ratios as a function of age.
Lynx cluster ETGs are the red dots, Lynx group ETGs are the blue triangles and CDF-S ETGs are the green stars.
For our sample, typical mass uncertainty is 40$\%$-60$\%$ and typical age uncertainty is 1-1.5 Gyrs.
We also report (squares with a black outline) data from literature for galaxies at $z \sim 1$-2 (\citet{bruzual07}: magenta, \citet{cimatti08}: light blue, \citet{muzzin09}: yellow; masses are normalized to a Salpeter IMF and ages are turned into star-formation weighted ages).
The M(MA05)/M(BC03) and M(CB07)/M(BC03) mass ratios decrease with age for ages lower than $\sim$1-1.5 Gyrs and then increase with age due to the effect of the TP-AGB phase modeling.
The mass ratio M(CB07)/M(MA05) decreases when age increases, however the explanation of this trend is less straightforward.\\}
	\label{fig:model_compar_mass_age}
\end{figure*}

	We compare in \textsc{Fig.} \ref{fig:model_compar_mass_age} the masses derived with the models.
Lynx cluster, Lynx group and CDF-S ETGs are in red dots, blue triangles and green stars respectively. 
To better understand those figures, we also report (squares with a black outline) data from  the literature for galaxies at $z \sim 1$-2 (\citet{bruzual07}: magenta; \citet{cimatti08}: light blue; \citet{muzzin09}: yellow).
The masses are normalized to a Salpeter IMF and ages are converted to star-formation weighted ages.
	
	The sample of \citet{bruzual07} is from \citet{daddi05} and has been studied by \citet{maraston06}.
It includes 7 ETGs ($1.4 \le z_{spec} \le 2.7$) selected to be passive with the $BzK$ criteria \citep{daddi04}.
The SED fitting ($V \to$ IRAC) for BC03 and MA05 models is made with various SFHs, a free metallicity and no dust.
For the CB07 models, the SED fitting uses a simple stellar population with a solar metallicity and no dust. 
The sample of \citet{cimatti08} is composed of 13 ($1.4 \le z_{spec} \le 2$, mainly early-type) galaxies selected in flux ($[$4.5$\mu$m$]$) and passively evolving.
The SED fitting ($B \to \lambda_{rest} \le 2.5 \mu$m) is made with exponentially declining SFHs, solar metallicity and dust (the authors note that the inclusion of dust has a weak impact on the derived ages and masses). 
The sample of \citet{muzzin09} includes 34 galaxies selected in flux ($K$) at $z \sim 2.3$ (about half of the sample has a spectroscopic redshift).
The SED fitting ($U \to$ IRAC + NIR spectrum) is made with exponentially declining SFHs, solar metallicity and dust.\\

	Though those data cannot be straightforwardly compared with ours, because the sample selection and the SED fitting procedure are different, their addition in the figures illustrates the trend seen with our data. 
When comparing BC03 with MA05/CB07, we observe that the mass ratio decreases with age for ages lower than $\sim$1-1.5 Gyrs and then increases with age.
This trend of mass ratio with age reflects the activity of the TP-AGB phase, which peaks at $\sim$1 Gyr.
In fact, the TP-AGB phase activity is important during the first Gyrs ($0.2 \lesssim$ t/Gyrs $\lesssim 2$ for a simple stellar population) with a peak around 1 Gyr \citep[e.g.,][]{maraston05}.
It is this peak shape of the TP-AGB phase activity that we observed on \textsc{Fig.} \ref{fig:model_compar_mass_age} when comparing BC03 with MA05/CB07, but spread on a longer timescale due to the extended star formation history.
As we approach the peak of the TP-AGB, MA05/CB07 model galaxies are redder and brighter at parity of mass with BC03.
This means that brighter galaxies are fitted by models with lower mass, up to half of the mass given by BC03.
We notice that the minimum of the mass ratio M(MA05)/M(BC03) is reached at $\sim$1 Gyr whereas the one of the mass ratio M(CB07)/M(BC03) is reached slightly later at $\sim$ 1.3 Gyrs.
When comparing MA05 masses with CB07 masses, we also see a trend with the age (the mass ratio M(CB07)/M(MA05) decreases when age increases), however the explanation of this trend is less straightforward.

\begin{figure*}
	   \includegraphics[width=\linewidth]{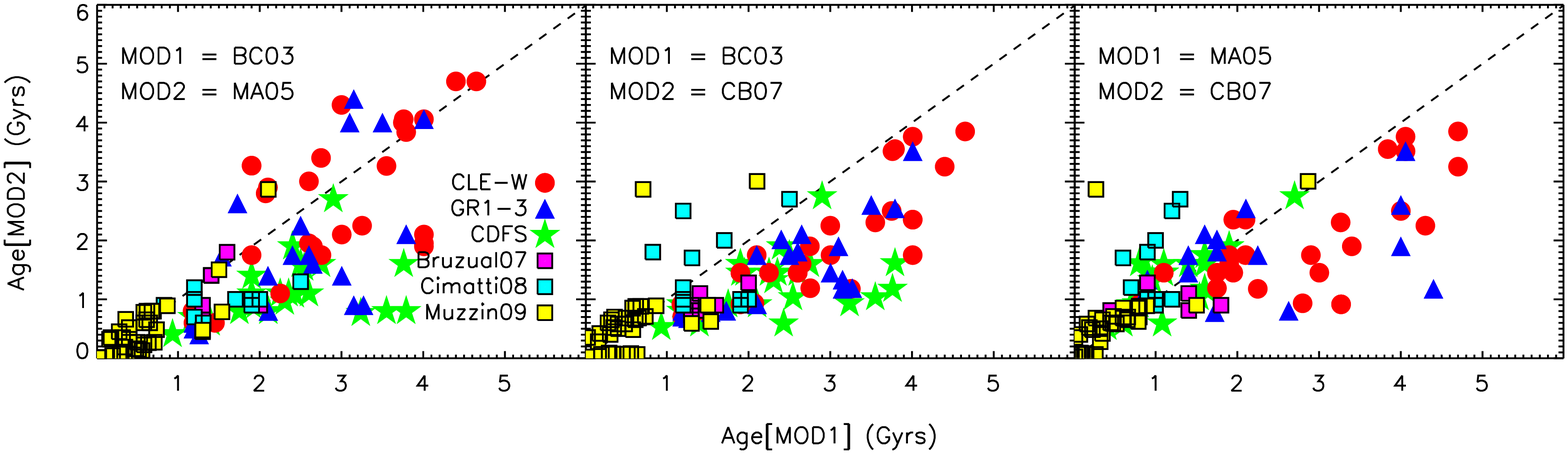}\\
	   \includegraphics[width=\linewidth]{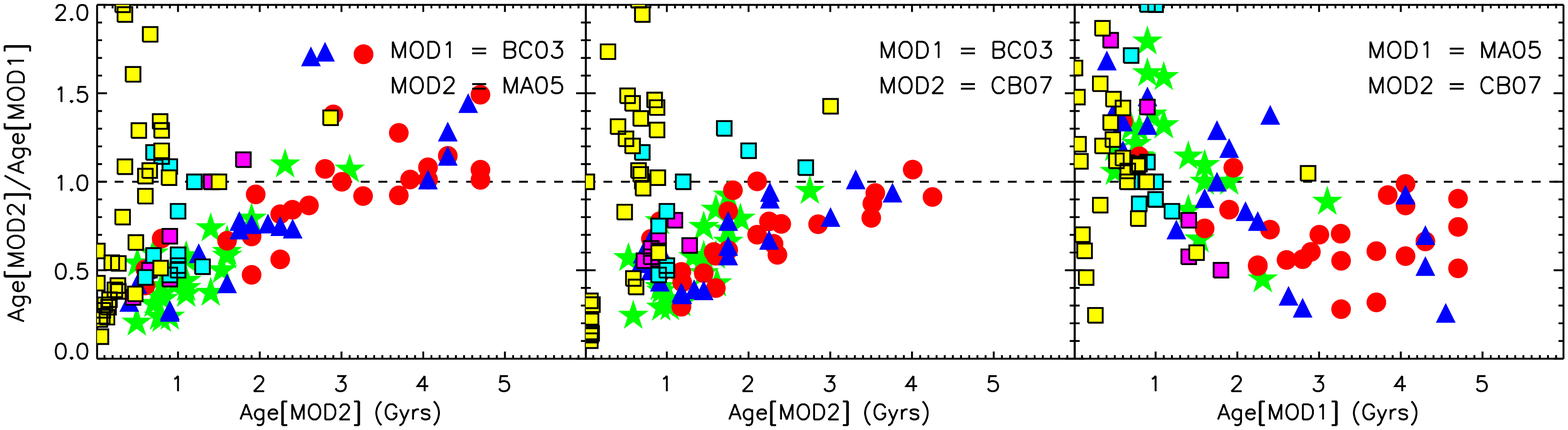}
	\caption{
The dependence of the estimated age on the stellar population models (BC03/MA05/CB07).
\textit{Top}: the age estimated with one model against the mass estimated with another.
\textit{Bottom}:  the age ratios as a function of age.
Symbols are the as in \textsc{Fig}.\ref{fig:model_compar_mass_age}.
For our sample, typical age uncertainty is 1-1.5 Gyrs.
We observe similar trends as in \textsc{Fig}.\ref{fig:model_compar_mass_age} due to the effect of the TP-AGB phase.
This trend is though more noisy, because of larger typical uncertainties.\\}
	\label{fig:model_compar_age_age}
\end{figure*}

	Concerning the age (\textsc{Fig.} \ref{fig:model_compar_age_age}), we observe similar but noisier trends, because of larger typical uncertainties.
The Age(MA05)/Age(BC03) and Age(CB07)/Age(BC03) ratios decrease with increasing age until $\sim$1 Gyr and then increase with age; the Age(CB07)/Age(MA05) ratio decreases with increasing age.
Our data do not allow us to determine the ages with precision (our typical uncertainty on age is $\sim$1-1.5 Gyrs). 

% FIGURE 5

% FIGURE 6

	While previous works \citep[e.g.,][]{bruzual07,cimatti08,muzzin09} already pointed out that estimated ages and masses depend strongly on the chosen stellar population model, our sample allows for a better insight into its dependence on the age, which was less clear in the three previous studies because the sample was too small or too homogeneous in age.

	The photometric catalog and the fitted parameters catalog are given in the appendix.	
	
%------------------------------------------------------------------------------------
% INFLUENCE OF THE MODELS AND FIT PARAMETERS: Aperture correction
%------------------------------------------------------------------------------------

\subsection{Uncertainties due to fixed fit parameters \label{sec:influence_fit_others}}

	We now estimate the uncertainty on the derived ages and masses due to the parameters we have kept fixed (aperture correction, IMF, dust).
The mean effect of each parameter are summarized in \textsc{Tab.}\ref{tab:fit_param_influence}.
We take as the reference photometry that which has an aperture correction using the PSF growth curve and a fit with a Salpeter IMF and no dust.
We report the mean and standard deviation of the ratio Age/Age$_{ref}$ for the ages and of log(M/M$_{ref}$) for the masses.

	Choosing an aperture correction with a a mean simulated galaxies growth curve instead of the PSF growth curve leads to a constant shift in mass of about $0.08$ dex and no significant change in age.
This result is expected: as the magnitude difference between the two methods is roughly independent of the band, using the PSF growth curve for aperture correction reduces the flux in the same manner regardless of the band, thus leaving the colors unchanged.

%------------------------------------------------------------------------------------
% INFLUENCE OF THE MODELS AND FIT PARAMETERS: IMF
%------------------------------------------------------------------------------------

	Changing the Salpeter IMF to a \citet{chabrier03} (resp. \citet{kroupa01} for the MA05 models) one results in a constant shift in mass of about $-0.25$ dex (resp. $-0.21$ dex ) and no significant change in age.

%------------------------------------------------------------------------------------
% INFLUENCE OF THE MODELS AND FIT PARAMETERS: Dust
%------------------------------------------------------------------------------------

	We also checked the influence of dust extinction by adding dust as a free parameter using $0 \le E(B-V) \le 0.4$, following the \citet{cardelli89} law.
We obtain on average $E(B-V) \sim 0.05-0.1$, except for the CDF-S ETGs fitted with BC03 models where $E(B-V) \sim 0.3$.
Adding dust is another way for BC03 models to match red colors of young galaxies due to TP-AGB phase.
We thus conclude, in agreement with e.g., \citet{rettura06}, \citet{cimatti08}, that the assumption of a dust-free model for our ETGs is reasonable.
	
%{\bf Additional uncertainties are due to the use of photometric redshifts versus spectroscopic redshift. This is often an uncertainty associated to high redshift samples. Since our uncertainties on photometric redshifts are very similar to usual uncertainties photometric redshifts obtained with multi-wavelength samples in the optical and near-infrared (M11), we adopt a fixed additional uncertainty of 0.3~dex as suggested from Monte Carlo simulations in similar works \citep{bundy05}.}	

% Summary
\begin{deluxetable*}{l c c | c c | c c}
	\tabletypesize{\small}
%	\tablewidth{}
%	\tablenum{}
%	\tablecolumns{}
	\tablecaption{Variations of age and mass due to the choice of the different fit parameters (see text). \label{tab:fit_param_influence}}
	\tablehead{
		\colhead{} 		& \multicolumn{2}{c}{BC03} 		& \multicolumn{2}{c}{MA05} 			& \multicolumn{2}{c}{CB07} \\
		\colhead{Parameter}& \colhead{Age}& \colhead{Mass}	& \colhead{Age}	& \colhead{Mass}	& \colhead{Age	}	& \colhead{Mass}\\
		\colhead{}			& \colhead{}	 & \colhead{(dex)}	& \colhead{}	 	& \colhead{(dex)}	& \colhead{} 		& \colhead{(dex)}
	}
	\startdata
	Aperture correction		& 1.06 $\pm$ 0.15	& \phantom{-}0.06 $\pm$ 0.04	& 1.11 $\pm$ 0.30	& \phantom{-}0.07 $\pm$ 0.07	& 1.06 $\pm$ 0.19	& \phantom{-}0.06 $\pm$ 0.07\\
	IMF\tablenotemark{a}	& 1.01 $\pm$ 0.10	& -0.25 $\pm$ 0.03	& 0.98 $\pm$ 0.13	& -0.21 $\pm$ 0.05	& 1.04 $\pm$ 0.15	& -0.24 $\pm$ 0.05\\
	Dust			 		& 0.57 $\pm$ 0.30	& -0.11 $\pm$ 0.11	& 0.82 $\pm$ 0.35	& -0.07 $\pm$ 0.13	& 0.75 $\pm$ 0.25	& -0.03 $\pm$ 0.10\\
	\enddata
	\tablenotetext{a}{For BC03 \& CB07 models: Chabrier IMF; for MA05 models: Kroupa IMF.}
\end{deluxetable*}

%@@@@@@@@@@@@@@@@@@@@@@@@@@@@@@@@@@@@@@@@@@@@@@@@@@@@@@@@@@@@@@@@@@@@@@@@
% RESULTS AND DISCUSSION
%@@@@@@@@@@@@@@@@@@@@@@@@@@@@@@@@@@@@@@@@@@@@@@@@@@@@@@@@@@@@@@@@@@@@@@@@

\section{Results \label{sec:results}}

	In this section, we present an analysis of the ages and masses estimated with the three different stellar population models (BC03/MA05/CB07).
For each figure of \textsc{Fig}.\ref{fig:color_plot}-\ref{fig:mass_age}, we present three figures corresponding to the three models, thus allowing a direct comparison of the dependence of the results on the models.

%------------------------------------------------------------------------------------
% RESULTS AND DISCUSSION: Color-color plot
%------------------------------------------------------------------------------------
\subsection{Color-color diagram \label{sec:color_color}}

\begin{figure*}
	   \includegraphics[width=\linewidth]{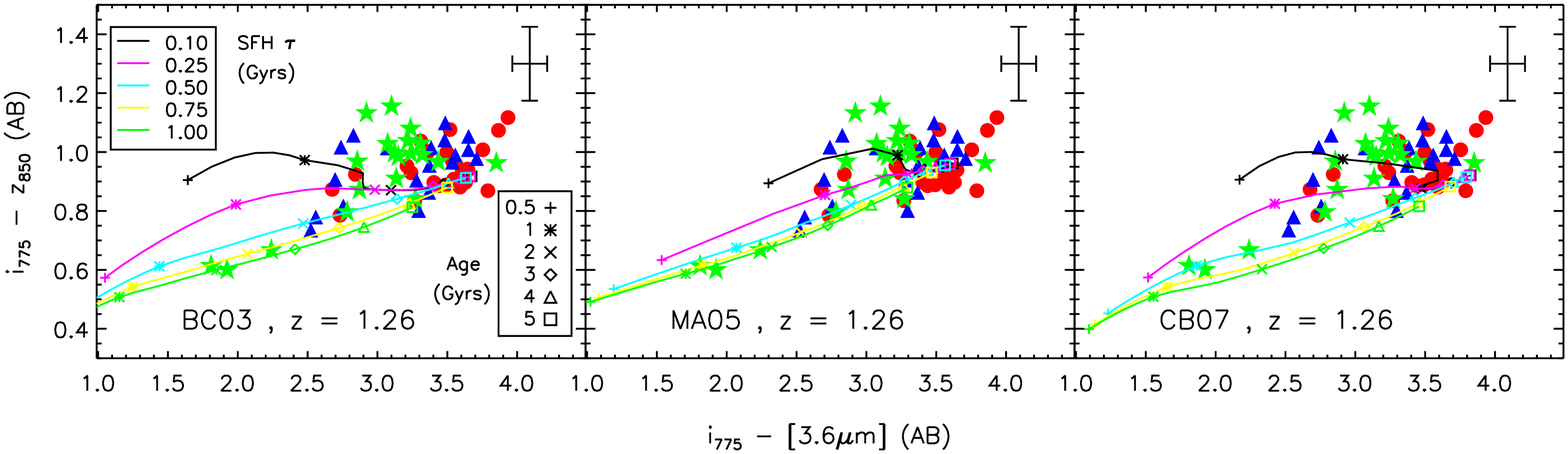}
	\caption{Color-color diagram:
Lynx cluster, Lynx group and CDF-S ETGs are in red dots, blue triangles and green stars respectively.
The models are represented by lines of different colors for different SFH $\tau$ ($0.1 \le \tau$ (Gyrs) $\le 1$) and for ages between 0.5 and 5 Gyrs.
The error bars in the upper right corner shows typical color uncertainties.\\}
	\label{fig:color_plot}
\end{figure*}

	In \textsc{Fig}.\ref{fig:color_plot} we compare observed galaxy colors to model predictions. 
We plot ($i_{775} -[3.6\mu$m$])$ against ($i_{775}-z_{850}$), where age and SFH separate better for our sample.
Lynx cluster ETGs are in red dots, Lynx group ETGs are in blue triangles and CDF-S ETGs are in green stars.
The models are represented by lines of different colors for different SFH $\tau$ ($0.1 \le \tau$ (Gyrs) $\le 1$) and for ages between 0.5 and 5 Gyrs.

% FIGURE 7
	
	Lynx cluster ETGs are less scattered than Lynx group and CDF-S ETGs and occupy a region corresponding to greater ages and shorter SFH $\tau$ for all models.
Lynx group and CDF-S ETGs show ages between 1 Gyr and 5 Gyrs for BC03 and 0.5 Gyr and 5 Gyrs for MA05 and BC07.
Again this is due to the different modeling of the TP-AGB phase.
For the range in ($i_{775} -[3.6\mu$m$])$ color between approximately 2.5 and 3.5, observations are reproduced by BC03 with ages on average between 0.5 and 1 Gyrs older than MA05/CB07.
We find that MA05 and CB07 predict colors that are closer to the observations.

	Given the large uncertainties in our SFH $\tau$ estimates, for an accurate analysis see our companion paper R11.

%------------------------------------------------------------------------------------
% RESULTS AND DISCUSSION: Relation between mass and NIR light
%------------------------------------------------------------------------------------

\subsection{Stellar mass and NIR rest-frame light \label{sec:ch1_mass}}

\begin{figure*}
	   \includegraphics[width=\linewidth]{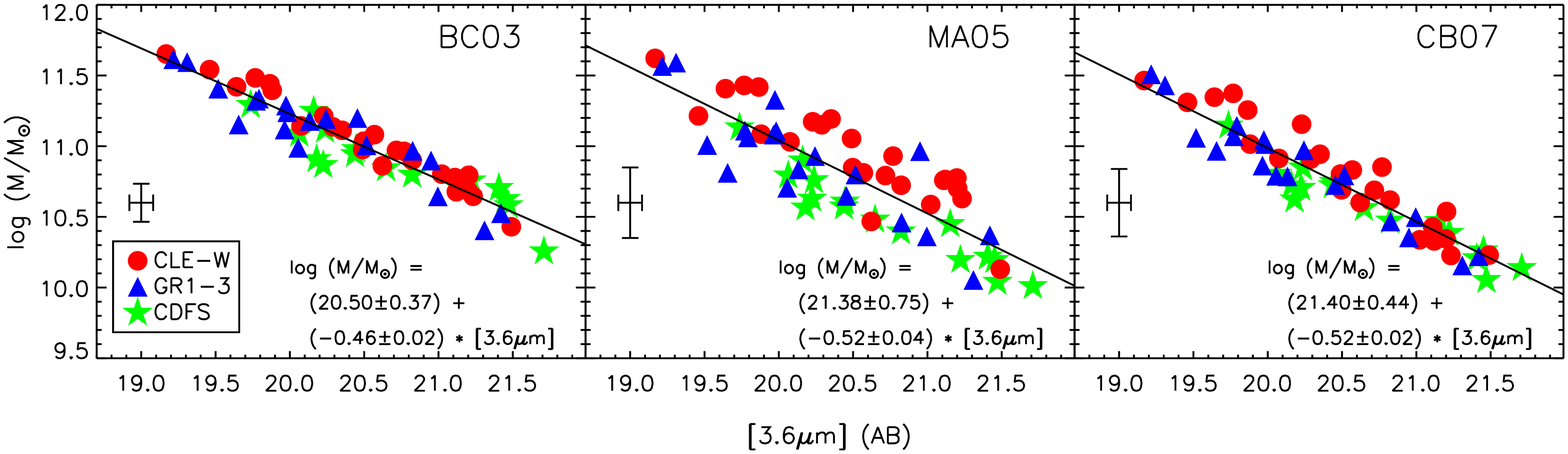}
	\caption{Relation between $[$3.6$\mu$m$] $ magnitude and the stellar mass:
the symbols are the same as in \textsc{Fig}.\ref{fig:color_plot}.
The solid line is the best linear fit given at the bottom of the plots.\\}
	\label{fig:ch1_mass}
\end{figure*}

	It is well-known that rest-frame the NIR magnitude correlates with galaxy stellar mass \citep[e.g.,][]{gavazzi93,kauffmann98}.
In \textsc{Fig}.\ref{fig:ch1_mass} we show our estimated mass versus the $[3.6\mu$m$]$ magnitude.
The symbols are the same as in \textsc{Fig}.\ref{fig:color_plot} and the solid line is the best linear fit given at the bottom of the plots.

% FIGURE 8

	At $z \sim 1.26$, the $[3.6\mu$m$]$ band probes rest-frame $H$  where the emission from old stars peaks.
We observe that the dispersion in this relation is smaller using BC03 models.
MA05 and CB07 show larger dispersion because of the different modeling of the NIR emission from the TP-AGB phase.
To quantify this dispersion, we measure the 1$\sigma$ dispersion of the log($M_{fit}/M$) distribution, where $M$ is the mass of the galaxy and $M_{fit}$ the mass corresponding to the linear fit.
We find a 1$\sigma$ dispersion of 0.10 for BC03 models, 0.22 for MA05 models and 0.13 for CB07 models.
For high-redshift galaxies, this effect should be taken into account when using NIR rest-frame magnitude as a proxy for the galaxy stellar mass \citep[see also][]{van-der-wel06}.

%------------------------------------------------------------------------------------
% RESULTS AND DISCUSSION: Masses and ages
%------------------------------------------------------------------------------------

\subsection{Ages and masses}

\begin{figure*}
   \includegraphics[width=\linewidth]{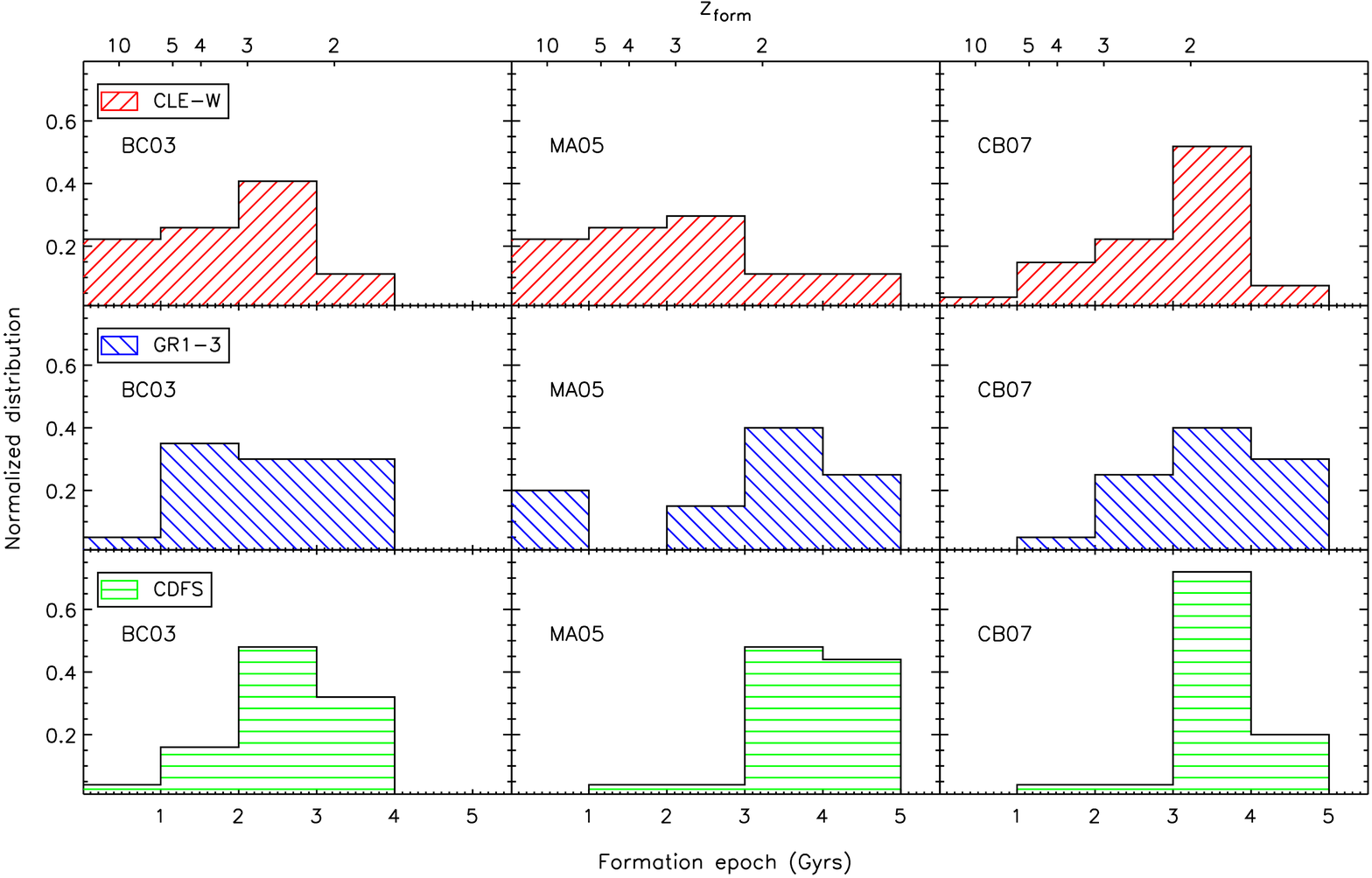}
	\caption{Normalized distribution of the formation epochs:
Lynx cluster ETGs (\textit{upper}, red tilted lines), Lynx group ETGs (\textit{middle}, blue tilted lines) and CDF-S ETGs (\textit{lower}, green horizontal lines).\\}
	\label{fig:hist_age}
\end{figure*}

\begin{figure*}
	 \includegraphics[width=\linewidth]{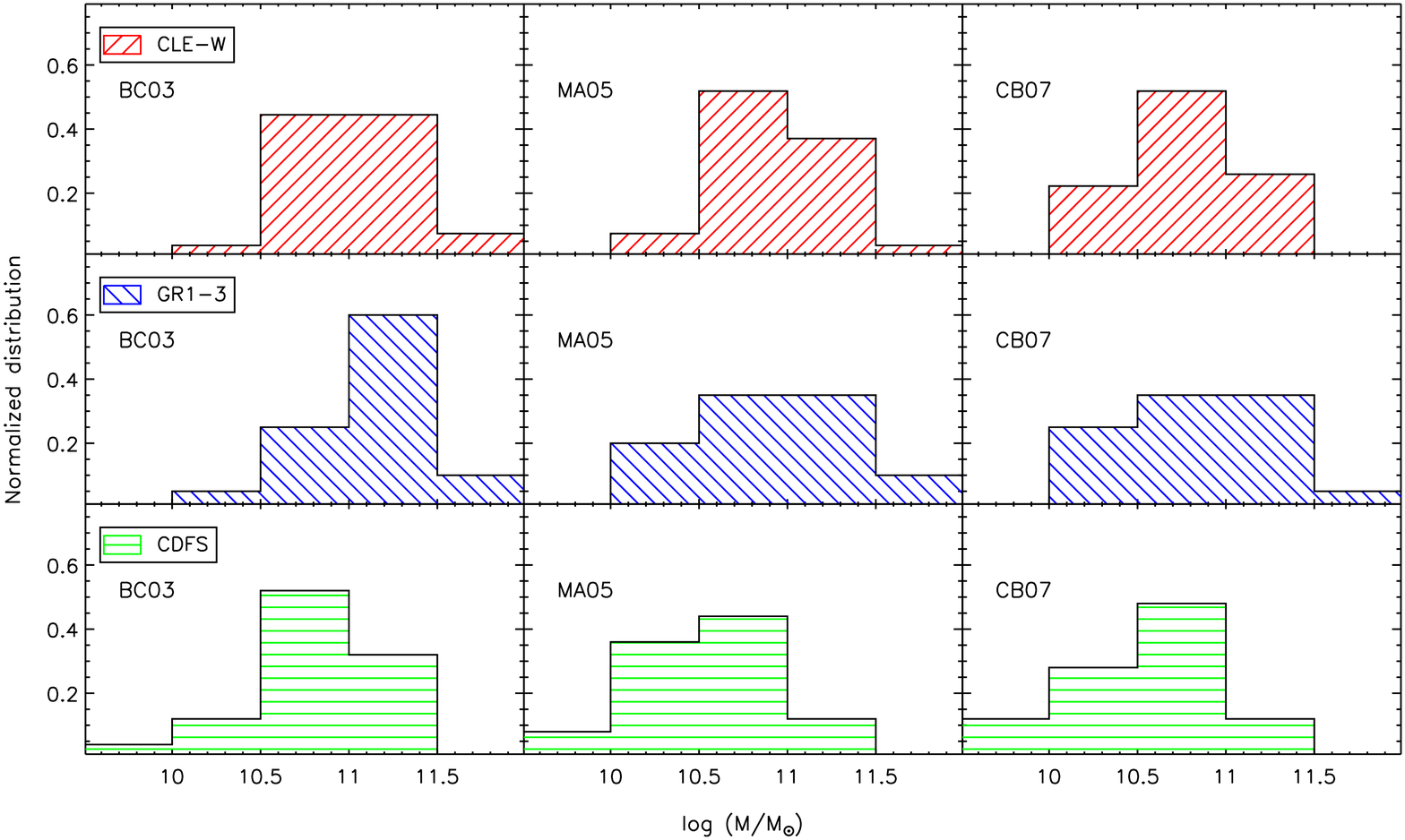}
	\caption{Normalized distribution of the masses:
Lynx cluster ETGs (\textit{upper}, red tilted lines), Lynx group ETGs (\textit{middle}, blue tilted lines) and CDF-S ETGs (\textit{lower}, green horizontal lines).\\}	
	\label{fig:hist_mass}
\end{figure*}

	We now discuss the estimated ages and masses. 
We plot in \textsc{Fig.}\ref{fig:hist_age} the normalized distribution of the formation epochs for Lynx cluster ETGs (red tilted lines), Lynx group ETGs (blue tilted lines) and CDF-S ETGs (green horizontal lines) and in \textsc{Fig.}\ref{fig:hist_mass} the normalized distribution of the masses.
For \textsc{Fig.}\ref{fig:hist_age}, we consider formation epoch and not age, because CDF-S ETGs are observed over a period spanning 0.85 Gyrs ($1.1 \lesssim z \lesssim 1.4$).
The formation epoch is obtained by subtracting the derived age to the age of the Universe at the observation redshift ($\sim$ 5 Gyrs).

% Age histogram
% FIGURE 9

	Concerning the formation epochs (\textsc{Fig.}\ref{fig:hist_age}), we find significant discrepancies between the model predictions.
The most significant one is for the CDF-S ETGs for MA05 and CB07 models on the one hand and BC03 models on the other hand: MA05 and CB07 models give similar formation epochs whereas BC03 models estimate a formation epoch on average $\sim$1 Gyr earlier.
Because of the lack of modeling of the TP-AGB phase, BC03 models artificially increase the age and the mass of the ETGs.
Ages and stellar masses derived with MA05 and CB07 seem to be more reliable.

	We note that the mean formation epoch estimated with MA05 models for Lynx cluster ETGs is about 0.7 Gyr earlier than the one estimated with CB07 models, while the mean formation epoch for Lynx group and CDF-S ETGs are similar when using MA05 or CB07 models.
	
	To quantify the difference in distribution for MA05 (resp. CB07) formation epochs in clusters and groups, we perform a Kolmogorov-Smirnov test and obtain that the hypothesis that the cluster and group formation epoch distributions are drawn from the same distribution can be rejected at a 2.5$\sigma$ significance level: there is a hint that the formation epoch distributions are different in the clusters and in the groups for MA05 and CB07 models.

%Furthermore, we also note that, when using MA05 or CB07 models, the mean formation epoch is consistently earlier for ETGs which lie in denser environments.

% Mass histogram
% FIGURE 10

	Concerning the masses (\textsc{Fig.}\ref{fig:hist_mass}), we first observe that, for the three models, the distribution and the mean value of the masses are similar for Lynx cluster and group ETGs.
Also, there are significantly less massive ETGs in the field than in the groups or the clusters (this is not the case for BC03 models that artificially increase masses for CDF-S ETGs). This is most likely due to the low probability of finding massive ETGs in the field and is consistent with other works \citep[e.g.,][]{fontana04,fontana06,scodeggio09,bolzonella10}.

	In the case of the mass distributions, a  Kolmogorov-Smirnov test does not permit to reject the hypothesis that the two distributions are different.
The probability that they are drawn from the same distribution is of 89\%  (MA05) and 46\% (CB07).
The masses are then most likely drawn from the same distribution.

	When using MA05 or CB07 models, the average formation epoch in clusters is consistently earlier than that in the groups and in the field.
In the field, this might be explained by the lack of massive ETGs: galaxies in the field are in average younger because they also are in average less massive.
However, in the groups galaxies are in average younger than in the clusters even if their mass distribution are most likely drawn from the same distribution.

\begin{figure*}
 \includegraphics[width=\linewidth]{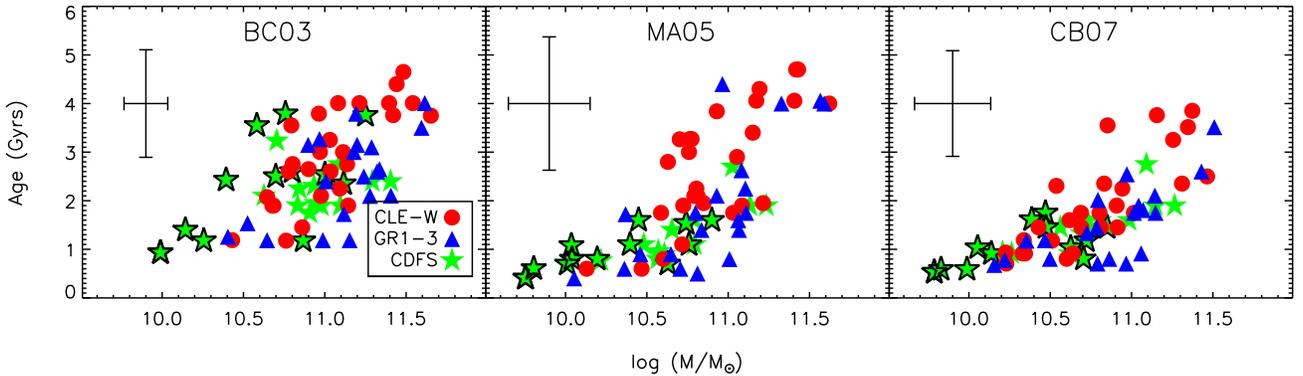}
	\caption{Age as a function of the mass:
the symbols are the same as in \textsc{Fig}.\ref{fig:color_plot}.
We mark with a black outline CDF-S ETGs with emission lines \citep{santini09}.
ETGs in clusters show a larger old and massive population than those in the groups and in the field.\\}
	\label{fig:mass_age}
\end{figure*}

	In \textsc{Fig.}\ref{fig:mass_age}, we plot the age as a function of the mass, with the same symbols as in \textsc{Fig.}\ref{fig:color_plot}.
We mark with a black outline CDF-S ETGs with emission lines \citep{santini09}.
We observe that, regardless of the model and of the environment, age increases with mass, which is consistent with the \textit{downsizing} scenario \citep{cowie96}.
According to this scenario, the most massive galaxies have formed first at high redshifts.
This scenario has long seemed to be in disagreement with the hierarchical scenario in which massive galaxies assemble their mass gradually.
Recently, \citet{de-lucia06} have shown that these two scenarios are in agreement if we distinguish the formation epoch of the stars and the assembly epoch of the galaxy. Old stellar populations in massive galaxies may be assembled by merging of less massive galaxies that were already dominated by old populations.
If however, the CDF-S sample lacks of low-mass/passive galaxies  (see discussion of the sample selection in $\S$\ref{sec:sample_sel}) because of our selection (as it can be seen in \textsc{Fig.}\ref{fig:mass_age}, the low-mass CDF-S ETGs mostly present emission lines ), we might miss this population in our analysis.
When using MA05 and CB07 models, Lynx cluster and group ETGs cover the entire range while the CDF-S sample shows a lack of  old ($\gtrsim 3$ Gyrs) or massive ($\gtrsim 10^{11} M_\odot$) ETGs (see above).

% Mass vs. age	
% FIGURE 11

%@@@@@@@@@@@@@@@@@@@@@@@@@@@@@@@@@@@@@@@@@@@@@@@@@@@@@@@@@@@@@@@@@@@@@@@@
% DISCUSSION AND CONCLUSION
%@@@@@@@@@@@@@@@@@@@@@@@@@@@@@@@@@@@@@@@@@@@@@@@@@@@@@@@@@@@@@@@@@@@@@@@@

\section{Discussion and conclusion \label{sec:conclusion}}

	In this work, we have studied a sample of \textcolor{red}{79} red ETGs spanning three different environments (cluster, group and field) at $z \sim 1.3$, combining observations of one of the most distant superclusters, the Lynx supercluster, and of the GOODS/CDF-S field.
We built a photometric catalog using 1.5\arcsec~ radius aperture photometry with PSF corrections, after exploring the possibility of using a growth curve which is closer to the real ETG growth curve.

	We then estimated the galaxy ages and stellar masses through SED fitting using different stellar population models (BC03, MA05 and CB07).
We show that the mass ratio between the masses estimated with different models depends on age.
When comparing MA05 and CB07 with BC03, we observe that their mass ratios M(MA05)/M(BC03) and M(CB07)/M(BC03) decrease with increasing age until an age of $\sim$1 Gyr and then increase.
This shape is due to inadequate modeling of the TP-AGB stellar phase activity in BC03.
Due to this problem, BC03 models artificially increase the age and the mass to fit the redder and more luminous emission of the TP-AGB phase.
When comparing CB07 with MA05, we also see a variation of the mass ratio with the age (the ratio M(CB07)/M(MA05) decreases with increasing age) but its explanation is less clear.
Concerning the age ratio, we see similar trends with age, though the dispersion is greater.
The already published mass and age estimations on similar ETGs at similar redshifts \citep{maraston06,bruzual07,cimatti08,muzzin09} with those three models are in agreement with this analysis.
The advantage of our sample when compared to previous results is the larger range in age, that permits us to identify parameter variations as a function of age.

	This means that current stellar population models give uncertain predictions.
In the next years it will be essential to compare their predictions to local and high redshift observations to understand their limitations in our interpretation of the data.
When interpreting observations at $z \sim 1.3$, while some results are stable and independent of the model, others depend significantly on the different modeling of the TP-AGB phase. 

This inability of the current stellar population models to consistently interpret the observations has been discussed in various previous works \citep[e.g.,][]{maraston06, cimatti08, eminian08,longhetti09, chen10,conroy10a,kelson10,kriek10}.
\citet{conroy10a} have compared BC03, MA05 and their own stellar population model to local observations and concluded that they cannot reproduce star cluster colors in the nearby Universe nor the properties of red sequence massive galaxies.
Both those authors  and \citet{kriek10} found that BC03 models better reproduce observed SEDs for post-starburst galaxies in local and high redshift observations, respectively. 

	Our sample at $z \sim 1.3$ shows that ETG colors and SED fits are slightly better  with the new models by M05 and CB07 when compared to BC03 \citep[e.g.,][]{maraston06,eminian08}, with BC03 predicted colors being bluer than the observations \citep[see also][]{mei06,mei09}.
All models though are consistent with color-color diagrams within the uncertainties and give reasonable SED fits.
What changes is mainly the parameters given by these fits and the observation interpretation, consistent with what is expected from the new implementations of the TP-AGB phase.
The advantage of our sample is the greater coverage in galaxy ages that permits us to identify the dependence of the mass and age ratios on age, and consequently on the weight of this new implementation on these parameters as a function of estimated galaxy age.

	Keeping in mind the dependences on the models, we obtain the following results:
\begin{enumerate}
\item Independent of the stellar population model and the environment, the most massive ETGs show consistently older ages.
\item According to the MA05 and CB07 models, the mass distribution is similar in the clusters and in the groups, whereas in our field sample there is a deficit of massive ($M \gtrsim 10^{11} M_\odot$) ETGs. This lack of massive ETGs in the field , firmly established in the local Universe \citep[e.g.,][]{bell03,kauffmann04}, is in agreement at high-redshift with recent studies in larger samples \citep[e.g.,][]{fontana04,fontana06,scodeggio09,bolzonella10}.
\item When using MA05 and CB07 models, although the mass distribution is similar in clusters and groups, on average, ages in groups are slightly younger ($\sim0.5$ Gyrs). Our field sample population is on average slightly younger than that in the clusters and groups, consistent with and because of the lack of massive ETGs. This small difference is less significant than the age difference for galaxies of different mass (see \textsc{Fig}.\ref{fig:mass_age}). These results are consistent with previous works at $ z \sim 1$ \citep{clemens09,cooper10}. Recent results from \citet{moresco10} \citep[see also][for similar results in the local Universe]{thomas10} also show a stronger dependence of galaxy age on mass and a much less significant dependence on environment. \citet{moresco10} for example find a difference in age of $<0.2$ Gyrs, that is consistent with our results considering the uncertainties on age estimates and that their sample probes less dense environments than ours (e.g. does not include massive clusters at $ z \sim 1$). 
\end{enumerate}

	In a $\Lambda$CDM cosmological model, galaxies have formed first in high density regions and assembled over time along filaments to build up larger structures such as galaxy groups and then clusters.
This cluster assembly is predicted to happen  mainly between $1 \lesssim z \lesssim 2$.
Our results at  $z \sim 1.3$ show that cluster and group galaxies have already formed the bulk of their massive ETG population and have a similar mass distribution, while in the field we do not observe massive ETGs. 
This might be due to the fact that in a $\Lambda$CDM, massive halos form in high density regions and/or with a different stellar formation history and stellar mass assembly \citep[e.g.,][]{poggianti06,poggianti10}. 
Regardless of the model and of the environment, the more massive galaxies are also the oldest, which is consistent with the \textit{downsizing} scenario \citep{cowie96}, in which the most massive galaxies have formed first.
This scenario can be reconciled with the $\Lambda$CDM  hierarchical scenario in which massive galaxies assemble their mass from the merging of less massive galaxies
\citep{de-lucia06} if  the stellar populations form first in less massive progenitors and then assemble later to form massive galaxies.

\acknowledgments
ACS was developed under NASA contract NAS 5-32865. This research 
has been supported by the NASA HST grant GO-10574.01-A, and Spitzer
grant for program 20694.
The {Space Telescope Science
Institute} is operated by AURA Inc., under NASA contract NAS5-26555.
Some of the data presented herein were obtained at the W.M. Keck
Observatory, which is operated as a scientific partnership among the
California Institute of Technology, the University of California and
the National Aeronautics and Space Administration. The Observatory was
made possible by the generous financial support of the W.M. Keck
Foundation.  
We thank Raphael Gobat and Veronica Strazzullo for useful discussions.

{\it Facilities:} \facility{HST (ACS)}, \facility{Spitzer (IRAC)}, \facility{KPNO:2.1m (FLAMINGOS)}, \facility{PO:4.5m (COSMIC)}, \facility{Keck:I (LRIS)}

\newpage
\appendix
\section{Lynx catalogs \label{sec:cat}}

%@@@@@@@@@@@@@@@@@@@@@@@@@@@@@@@@@@@@@@@@@@@@@@@@@@@@@@@@@@@@@@@@@@@@@@@@
% APPENDIX A:  PHOTOMETRIC CATALOG
%@@@@@@@@@@@@@@@@@@@@@@@@@@@@@@@@@@@@@@@@@@@@@@@@@@@@@@@@@@@@@@@@@@@@@@@@

%\section{Lynx ETG magnitudes \label{sec:cat_phot}}

\begin{deluxetable}{l c c c c c c c c c}
\setlength{\tabcolsep}{0.05in}
\tabletypesize{\scriptsize}
%	\rotate
%	\tablewidth{}
%	\tablenum{}
	\tablecolumns{10}
\tablecaption{Lynx cluster ETG astrometry and magnitudes \label{tab:cat_phot_cl}}
\tablehead{
\colhead{ID}	&
\colhead{R.A.} &
\colhead{DEC.} &
\colhead{\textit{R}}		&
\colhead{\textit{i$_{775}$}}	&
\colhead{\textit{z$_{850}$}}	&
\colhead{\textit{J}}		&
\colhead{\textit{K$_s$}}	&
\colhead{\textit{$[$3.6$\mu$m$] $}}	&	
\colhead{\textit{$[$4.5$\mu$m$] $}}\\
\colhead{}	&
\colhead{(J2000)} &
\colhead{(J2000)} &
\colhead{(AB mag)}		&
\colhead{(AB mag)}	&
\colhead{(AB mag)}	&
\colhead{(AB mag)}		&
\colhead{(AB mag)}	&
\colhead{(AB mag)}	&	
\colhead{(AB mag)}
}
\startdata
\multicolumn{10}{c}{\textsc{Lynx Cluster E ($\langle z \rangle = 1.261$)}} \\
\cline{1-10}\\
%\cutinhead{\textsc{Lynx Cluster E ($\langle z \rangle = 1.261$)}}
4945 & 08 48 49.99 & +44 52 01.78 & 24.81 $\pm$ 0.11 & 23.68 $\pm$ 0.09 & 22.56 $\pm$ 0.07 & 21.21 $\pm$ 0.13 & 20.28 $\pm$ 0.10 & 19.77 $\pm$ 0.08 & 19.87 $\pm$ 0.08 \\
6229 & 08 48 55.90 & +44 51 54.99 & 25.58 $\pm$ 0.17 & 24.30 $\pm$ 0.12 & 23.30 $\pm$ 0.09 & U & 21.42 $\pm$ 0.15 & 20.82 $\pm$ 0.08 & 20.93 $\pm$ 0.09 \\
6090 & 08 48 56.64 & +44 51 55.76 & 25.26 $\pm$ 0.14 & 24.20 $\pm$ 0.11 & 23.25 $\pm$ 0.09 & U & 21.27 $\pm$ 0.15 & 20.57 $\pm$ 0.08 & 20.82 $\pm$ 0.09 \\
5355 & 08 48 57.66 & +44 53 48.69 & 25.52 $\pm$ 0.17 & 24.39 $\pm$ 0.12 & 23.43 $\pm$ 0.10 & 21.95 $\pm$ 0.21 & 21.47 $\pm$ 0.18 & 21.20 $\pm$ 0.08 & 21.51 $\pm$ 0.10 \\
8713 & 08 48 57.85 & +44 50 55.32 & 24.38 $\pm$ 0.10 & 23.95 $\pm$ 0.10 & 23.16 $\pm$ 0.08 & U & 21.67 $\pm$ 0.22 & 21.23 $\pm$ 0.08 & 21.45 $\pm$ 0.10 \\
5817 & 08 48 57.91 & +44 51 52.25 & 25.21 $\pm$ 0.14 & 23.94 $\pm$ 0.10 & 22.99 $\pm$ 0.08 & 21.81 $\pm$ 0.21 & 21.02 $\pm$ 0.13 & 20.35 $\pm$ 0.08 & 20.43 $\pm$ 0.09 \\
5634 & 08 48 58.53 & +44 51 33.25 & 24.47 $\pm$ 0.10 & 23.40 $\pm$ 0.08 & 22.42 $\pm$ 0.07 & 21.30 $\pm$ 0.13 & 20.48 $\pm$ 0.10 & 19.88 $\pm$ 0.08 & 19.91 $\pm$ 0.08 \\
5693 & 08 48 58.60 & +44 51 57.21 & B & 22.92 $\pm$ 0.07 & 22.15 $\pm$ 0.07 & B & B & B & B \\
5680 & 08 48 58.63 & +44 51 59.46 & 24.89 $\pm$ 0.11 & 23.60 $\pm$ 0.08 & 22.84 $\pm$ 0.08 & B & B & B & B \\
5794 & 08 48 58.67 & +44 51 56.97 & B & 23.04 $\pm$ 0.07 & 22.18 $\pm$ 0.07 & B & B & B & B \\
8495 & 08 48 58.93 & +44 50 33.77 & 25.29 $\pm$ 0.14 & 24.03 $\pm$ 0.10 & 23.20 $\pm$ 0.09 & B & B & B & B \\
5748 & 08 48 58.95 & +44 52 10.90 & 24.71 $\pm$ 0.10 & 23.71 $\pm$ 0.09 & 22.64 $\pm$ 0.07 & 21.35 $\pm$ 0.13 & 20.33 $\pm$ 0.10 & 19.87 $\pm$ 0.08 & 19.91 $\pm$ 0.08 \\
5689 & 08 48 59.10 & +44 52 04.64 & 25.53 $\pm$ 0.17 & 24.47 $\pm$ 0.14 & 23.46 $\pm$ 0.10 & 22.09 $\pm$ 0.27 & 21.58 $\pm$ 0.18 & 21.12 $\pm$ 0.08 & 21.57 $\pm$ 0.10 \\
5876 & 08 48 59.72 & +44 52 51.28 & 24.16 $\pm$ 0.09 & 23.25 $\pm$ 0.08 & 22.35 $\pm$ 0.07 & 21.13 $\pm$ 0.12 & 20.24 $\pm$ 0.10 & 19.64 $\pm$ 0.08 & 19.85 $\pm$ 0.08 \\
5602 & 08 49 00.32 & +44 52 14.39 & 25.56 $\pm$ 0.17 & 23.71 $\pm$ 0.09 & 22.78 $\pm$ 0.07 & 21.84 $\pm$ 0.21 & 20.89 $\pm$ 0.13 & 20.49 $\pm$ 0.08 & 20.80 $\pm$ 0.09 \\
8662 & 08 49 01.07 & +44 52 09.65 & 24.79 $\pm$ 0.11 & 24.40 $\pm$ 0.12 & 23.46 $\pm$ 0.10 & U & 21.41 $\pm$ 0.15 & 20.77 $\pm$ 0.08 & 20.97 $\pm$ 0.09 \\
8041 & 08 49 01.52 & +44 50 49.73 & B & 23.60 $\pm$ 0.08 & 22.53 $\pm$ 0.07 & 21.79 $\pm$ 0.21 & 20.90 $\pm$ 0.13 & B & B \\
8625 & 08 49 03.31 & +44 53 04.12 & 25.15 $\pm$ 0.13 & 24.15 $\pm$ 0.11 & 23.27 $\pm$ 0.09 & U & 21.64 $\pm$ 0.18 & 21.49 $\pm$ 0.09 & 21.70 $\pm$ 0.10 \\
7653 & 08 49 04.52 & +44 50 16.42 & - & 24.41 $\pm$ 0.12 & 23.51 $\pm$ 0.10 & U & 21.77 $\pm$ 0.22 & 21.02 $\pm$ 0.08 & 21.29 $\pm$ 0.09 \\
8047 & 08 49 05.34 & +44 52 03.79 & 24.32 $\pm$ 0.09 & 23.45 $\pm$ 0.08 & 22.52 $\pm$ 0.07 & 21.53 $\pm$ 0.15 & 21.18 $\pm$ 0.14 & 20.62 $\pm$ 0.08 & 20.75 $\pm$ 0.09 \\
7475 & 08 49 05.96 & +44 50 37.00 & - & 23.65 $\pm$ 0.08 & 22.60 $\pm$ 0.07 & 21.87 $\pm$ 0.21 & 20.67 $\pm$ 0.10 & B & B \\
\cutinhead{\textsc{Lynx Cluster W ($\langle z \rangle = 1.273$)}}
1745 & 08 48 29.71 & +44 52 49.68 & 25.67 $\pm$ 0.20 & 24.53 $\pm$ 0.14 & 23.64 $\pm$ 0.11 & U & 21.74 $\pm$ 0.22 & 21.11 $\pm$ 0.08 & 21.32 $\pm$ 0.09 \\
1486 & 08 48 31.72 & +44 54 42.95 & 25.05 $\pm$ 0.13 & 24.45 $\pm$ 0.12 & 23.61 $\pm$ 0.11 & U & 21.76 $\pm$ 0.22 & 21.20 $\pm$ 0.08 & 21.54 $\pm$ 0.10 \\
1794 & 08 48 32.78 & +44 54 07.22 & 25.29 $\pm$ 0.14 & 24.28 $\pm$ 0.12 & 23.19 $\pm$ 0.09 & B & 21.13 $\pm$ 0.14 & B & B \\
1922 & 08 48 32.99 & +44 53 46.69 & 24.51 $\pm$ 0.10 & 23.37 $\pm$ 0.08 & 22.33 $\pm$ 0.07 & 21.12 $\pm$ 0.12 & 20.42 $\pm$ 0.10 & 20.08 $\pm$ 0.08 & 20.24 $\pm$ 0.08 \\
1525 & 08 48 33.01 & +44 55 11.92 & 25.02 $\pm$ 0.13 & 24.00 $\pm$ 0.10 & 22.92 $\pm$ 0.08 & 21.87 $\pm$ 0.21 & 20.87 $\pm$ 0.13 & 20.50 $\pm$ 0.08 & 20.72 $\pm$ 0.09 \\
1962 & 08 48 33.04 & +44 53 39.75 & 25.38 $\pm$ 0.14 & 24.19 $\pm$ 0.11 & 23.30 $\pm$ 0.09 & U & 21.23 $\pm$ 0.14 & 20.72 $\pm$ 0.08 & 20.92 $\pm$ 0.09 \\
2094 & 08 48 34.08 & +44 53 32.32 & 24.86 $\pm$ 0.11 & 24.00 $\pm$ 0.10 & 23.13 $\pm$ 0.08 & U & 20.94 $\pm$ 0.13 & 20.23 $\pm$ 0.08 & 20.24 $\pm$ 0.08 \\
2343 & 08 48 35.98 & +44 53 36.12 & 24.04 $\pm$ 0.09 & 22.90 $\pm$ 0.07 & 21.89 $\pm$ 0.07 & 20.64 $\pm$ 0.10 & 19.75 $\pm$ 0.08 & 19.17 $\pm$ 0.08 & 19.31 $\pm$ 0.08 \\
2195 & 08 48 36.17 & +44 54 17.30 & 24.08 $\pm$ 0.09 & 23.03 $\pm$ 0.07 & 22.15 $\pm$ 0.07 & 20.98 $\pm$ 0.12 & 20.08 $\pm$ 0.09 & 19.46 $\pm$ 0.08 & 19.63 $\pm$ 0.08 \\
2571 & 08 48 37.08 & +44 53 34.05 & 25.07 $\pm$ 0.13 & 23.82 $\pm$ 0.09 & 22.90 $\pm$ 0.08 & 21.52 $\pm$ 0.15 & 20.74 $\pm$ 0.10 & 20.29 $\pm$ 0.08 & 20.48 $\pm$ 0.09 \\
\enddata
\tablecomments{Magnitudes are 1.5" radius aperture magnitudes; aperture corrections are estimated with PSF growth curves normalized at 7" radius (cf. \textsc{Tab.} \ref{tab:apcor}); magnitude errors are estimated via simulations (cf. section \ref{sec:mag_err}); all magnitudes are in the AB system and are corrected for Galactic extinction. U: the galaxy is not detected in the image; B: the galaxy is blended with a close object; -: the galaxy is not on our images. 
If one wants to use aperture correction derived from simulated galaxies growth curve, one has to use the following relations: mag$_{GAL}$ = mag$_{PSF}$ - ApC$_{PSF}$ + ApC$_{GAL}$ and $\sigma_{GAL} = \sqrt{\sigma_{PSF}^2 - \sigma_{ApC,PSF}^2}$, where mag$_{PSF}$ and $\sigma_{PSF}$ denote the magnitudes and their uncertainty using the PSF growth curve for aperture correction (in this table), mag$_{GAL}$ and $\sigma_{GAL}$ the magnitudes and their uncertainty using the simulated galaxies growth curve for aperture correction, ApC$_*$ the aperture correction and $\sigma_{ApC,PSF}$ the uncertainty on the PSF aperture correction (all given \textsc{Tab}.\ref{tab:apcor}).}
\end{deluxetable}

\begin{deluxetable}{l c c c c c c c c c}
\setlength{\tabcolsep}{0.05in}
\tabletypesize{\scriptsize}
%	\rotate
%	\tablewidth{}
%	\tablenum{}
	\tablecolumns{10}
\tablecaption{Lynx group ETG astrometry and magnitudes \label{tab:cat_phot_gr}}
\tablehead{
\colhead{ID}	&
\colhead{R.A.} &
\colhead{DEC.} &
\colhead{\textit{R}}		&
\colhead{\textit{i$_{775}$}}	&
\colhead{\textit{z$_{850}$}}	&
\colhead{\textit{J}}		&
\colhead{\textit{K$_s$}}	&
\colhead{\textit{$[$3.6$\mu$m$] $}}	&	
\colhead{\textit{$[$4.5$\mu$m$] $}}\\
\colhead{}	&
\colhead{(J2000)} &
\colhead{(J2000)} &
\colhead{(AB mag)}		&
\colhead{(AB mag)}	&
\colhead{(AB mag)}	&
\colhead{(AB mag)}		&
\colhead{(AB mag)}	&
\colhead{(AB mag)}	&	
\colhead{(AB mag)}
}
\startdata
\cutinhead{\textsc{Lynx Group 1 ($\langle z \rangle = 1.266$)}}
518 & 08 49 03.52 & +44 53 21.62 & 24.49 $\pm$ 0.14 & 23.81 $\pm$ 0.09 & 23.07 $\pm$ 0.08 & U & 22.10 $\pm$ 0.33 & 21.31 $\pm$ 0.08 & 22.70 $\pm$ 0.15 \\
1339 & 08 49 08.32 & +44 53 48.32 & 23.92 $\pm$ 0.12 & 22.94 $\pm$ 0.07 & 21.88 $\pm$ 0.07 & 20.81 $\pm$ 0.11 & 19.68 $\pm$ 0.08 & 19.31 $\pm$ 0.08 & 19.31 $\pm$ 0.08 \\
1024 & 08 49 09.00 & +44 52 44.08 & 24.55 $\pm$ 0.15 & 23.72 $\pm$ 0.09 & 22.70 $\pm$ 0.07 & U & 21.22 $\pm$ 0.14 & 21.00 $\pm$ 0.08 & 21.02 $\pm$ 0.09 \\
825 & 08 49 11.24 & +44 51 29.19 & 23.41 $\pm$ 0.10 & 22.57 $\pm$ 0.07 & 21.55 $\pm$ 0.07 & 20.70 $\pm$ 0.10 & 19.78 $\pm$ 0.08 & 19.52 $\pm$ 0.08 & 19.51 $\pm$ 0.08 \\
1249 & 08 49 12.27 & +44 52 13.05 & B & 23.40 $\pm$ 0.08 & 22.41 $\pm$ 0.07 & B & B & B & B \\
1085 & 08 49 13.69 & +44 51 18.82 & - & 23.04 $\pm$ 0.07 & 22.09 $\pm$ 0.07 & 21.00 $\pm$ 0.12 & 20.17 $\pm$ 0.09 & - & - \\
\cutinhead{\textsc{Lynx Group 2 ($\langle z \rangle = 1.262$)}}
1636 & 08 49 00.92 & +44 58 49.15 & 24.45 $\pm$ 0.14 & 23.45 $\pm$ 0.08 & 22.35 $\pm$ 0.07 & 21.28 $\pm$ 0.13 & 20.41 $\pm$ 0.10 & 19.98 $\pm$ 0.08 & 20.12 $\pm$ 0.08 \\
939 & 08 49 03.43 & +44 56 38.59 & 23.64 $\pm$ 0.11 & 22.87 $\pm$ 0.07 & 21.80 $\pm$ 0.07 & 21.01 $\pm$ 0.12 & 20.49 $\pm$ 0.10 & 20.06 $\pm$ 0.08 & 20.28 $\pm$ 0.08 \\
1383 & 08 49 03.99 & +44 57 23.37 & 24.70 $\pm$ 0.15 & 23.65 $\pm$ 0.09 & 22.68 $\pm$ 0.07 & 22.16 $\pm$ 0.27 & 20.95 $\pm$ 0.13 & 20.13 $\pm$ 0.08 & 20.31 $\pm$ 0.08 \\
2000 & 08 49 07.15 & +44 57 52.04 & 23.78 $\pm$ 0.11 & 22.91 $\pm$ 0.07 & 21.93 $\pm$ 0.07 & 20.71 $\pm$ 0.10 & 19.83 $\pm$ 0.08 & 19.22 $\pm$ 0.08 & 19.36 $\pm$ 0.08 \\
2519 & 08 49 08.66 & +44 58 43.26 & 24.86 $\pm$ 0.18 & 24.09 $\pm$ 0.11 & 23.18 $\pm$ 0.08 & U & 21.38 $\pm$ 0.15 & 20.83 $\pm$ 0.08 & 20.84 $\pm$ 0.09 \\
1791 & 08 49 10.25 & +44 56 34.50 & 23.17 $\pm$ 0.10 & 22.33 $\pm$ 0.07 & 21.42 $\pm$ 0.07 & 20.72 $\pm$ 0.10 & 19.93 $\pm$ 0.09 & 19.66 $\pm$ 0.08 & 19.94 $\pm$ 0.08 \\
\cutinhead{\textsc{Lynx Group 3 ($\langle z \rangle = 1.264$)}}
137 & 08 48 53.26 & +44 44 22.39 & - & 23.82 $\pm$ 0.09 & 22.80 $\pm$ 0.08 & U & 21.03 $\pm$ 0.14 & 20.45 $\pm$ 0.08 & 19.89 $\pm$ 0.08 \\
542 & 08 48 55.14 & +44 44 58.83 & - & 23.26 $\pm$ 0.08 & 22.21 $\pm$ 0.07 & 20.97 $\pm$ 0.12 & 20.29 $\pm$ 0.10 & 19.79 $\pm$ 0.08 & 19.91 $\pm$ 0.08 \\
1135 & 08 48 56.28 & +44 46 45.62 & - & 23.13 $\pm$ 0.08 & 22.17 $\pm$ 0.07 & 21.10 $\pm$ 0.12 & 20.21 $\pm$ 0.09 & 19.77 $\pm$ 0.08 & 19.92 $\pm$ 0.08 \\
889 & 08 48 56.63 & +44 45 39.90 & - & 24.58 $\pm$ 0.14 & 23.57 $\pm$ 0.10 & U & 21.47 $\pm$ 0.18 & 20.95 $\pm$ 0.08 & B \\
1431 & 08 48 57.31 & +44 47 08.01 & - & 23.86 $\pm$ 0.10 & 23.00 $\pm$ 0.08 & 21.77 $\pm$ 0.18 & 21.05 $\pm$ 0.14 & 20.52 $\pm$ 0.08 & 20.71 $\pm$ 0.09 \\
1064 & 08 48 57.79 & +44 45 57.51 & - & 23.96 $\pm$ 0.10 & 23.18 $\pm$ 0.08 & U & 21.86 $\pm$ 0.26 & 21.42 $\pm$ 0.08 & 21.58 $\pm$ 0.10 \\
1136 & 08 48 57.96 & +44 46 04.53 & - & 23.51 $\pm$ 0.08 & 22.52 $\pm$ 0.07 & 21.45 $\pm$ 0.15 & 20.35 $\pm$ 0.10 & 19.97 $\pm$ 0.08 & 20.07 $\pm$ 0.08 \\
1775 & 08 49 01.62 & +44 46 28.23 & - & 23.52 $\pm$ 0.08 & 22.71 $\pm$ 0.07 & 21.54 $\pm$ 0.15 & 20.77 $\pm$ 0.10 & 20.24 $\pm$ 0.08 & 20.44 $\pm$ 0.09 \\
1731 & 08 49 04.43 & +44 45 08.65 & - & 22.70 $\pm$ 0.07 & 21.88 $\pm$ 0.07 & 21.19 $\pm$ 0.13 & 20.33 $\pm$ 0.10 & 19.96 $\pm$ 0.08 & 20.19 $\pm$ 0.08 \\
\enddata
\tablecomments{Cf. \textsc{Tab}.\ref{tab:cat_phot_cl}}
\end{deluxetable}

%@@@@@@@@@@@@@@@@@@@@@@@@@@@@@@@@@@@@@@@@@@@@@@@@@@@@@@@@@@@@@@@@@@@@@@@@
% APPENDIX B:  MASS/AGE CATALOG
%@@@@@@@@@@@@@@@@@@@@@@@@@@@@@@@@@@@@@@@@@@@@@@@@@@@@@@@@@@@@@@@@@@@@@@@@
	
%\newpage
%\section{Lynx ETG ages and masses \label{sec:cat_mass}}

\begin{deluxetable}{l c c c || c c c || c c c}
	\setlength{\tabcolsep}{0.15in}
	\tabletypesize{\scriptsize}
%	\rotate
%	\tablewidth{0pt}
%	\tablenum{}
	\tablecolumns{7}
	\tablecaption{Lynx cluster ETG stellar population ages and stellar masses \label{tab:cat_mass_cl}}
	\tablehead{
		\colhead{} 	& 
		\multicolumn{3}{c}{BC03} & 
		\multicolumn{3}{c}{MA05} & 
		\multicolumn{3}{c}{CB07} \\
		\multicolumn{10}{c}{} \\
		\colhead{ID} & 
		\colhead{Age} & 
		\colhead{log $\frac{M}{M_{\sun}}$} & 
		\colhead{SFH $\tau$} & 
		\colhead{Age} & 
		\colhead{log $\frac{M}{M_{\sun}}$} & 
		\colhead{SFH $\tau$} & 
		\colhead{Age} & 
		\colhead{log $\frac{M}{M_{\sun}}$} & 
		\colhead{SFH $\tau$} \\
		\colhead{} & 
		\colhead{(Gyrs)} & 
		\colhead{} & 
		\colhead{(Gyrs)} & 
		\colhead{(Gyrs)} & 
		\colhead{} & 
		\colhead{(Gyrs)} & 
		\colhead{(Gyrs)} &
		\colhead{} &
		\colhead{(Gyrs)}
	}
	\startdata
%\multicolumn{10}{c}{\textsc{Lynx Cluster E ($\langle z \rangle = 1.261$)}} \\
%\cline{1-10}\\
\cutinhead{\textsc{Lynx Cluster E ($\langle z \rangle = 1.261$)}}
4945 & 4.7$^{+0.3}_{-1.5}$ & 11.48$^{+0.04}_{-0.19}$ & 0.1$^{+0.4}_{-0.0}$ & 4.7$^{+0.2}_{-0.8}$ & 11.43$^{+0.04}_{-0.07}$ & 0.1$^{+0.4}_{-0.0}$ & 3.9$^{+0.5}_{-1.3}$ & 11.37$^{+0.11}_{-0.16}$ & 0.4$^{+0.1}_{-0.3}$ \\
6229 & 2.7$^{+2.0}_{-0.6}$ & 10.90$^{+0.16}_{-0.09}$ & 0.1$^{+0.7}_{-0.0}$ & 1.9$^{+2.8}_{-0.5}$ & 10.72$^{+0.33}_{-0.16}$ & 0.1$^{+0.7}_{-0.0}$ & 1.6$^{+1.6}_{-0.4}$ & 10.62$^{+0.25}_{-0.23}$ & 0.1$^{+0.4}_{-0.0}$ \\
6090 & 4.0$^{+0.3}_{-2.2}$ & 11.08$^{+0.07}_{-0.15}$ & 0.8$^{+0.3}_{-0.7}$ & 2.1$^{+2.6}_{-0.7}$ & 10.81$^{+0.34}_{-0.17}$ & 0.1$^{+0.7}_{-0.0}$ & 2.4$^{+1.8}_{-1.5}$ & 10.83$^{+0.18}_{-0.39}$ & 0.4$^{+0.4}_{-0.3}$ \\
5355 & 1.9$^{+1.3}_{-0.7}$ & 10.68$^{+0.11}_{-0.24}$ & 0.1$^{+0.4}_{-0.0}$ & 3.3$^{+0.8}_{-2.8}$ & 10.77$^{+0.08}_{-0.43}$ & 0.8$^{+0.3}_{-0.7}$ & 0.9$^{+1.5}_{-0.1}$ & 10.35$^{+0.23}_{-0.08}$ & 0.1$^{+0.4}_{-0.0}$ \\
8713 & 2.1$^{+2.0}_{-0.8}$ & 10.65$^{+0.18}_{-0.10}$ & 0.8$^{+0.8}_{-0.3}$ & 2.8$^{+0.8}_{-3.1}$ & 10.63$^{+0.09}_{-0.50}$ & 1.5$^{+0.5}_{-1.3}$ & 0.9$^{+3.5}_{-0.3}$ & 10.23$^{+0.45}_{-0.02}$ & 0.4$^{+1.6}_{-0.1}$ \\
5817 & 3.0$^{+1.5}_{-0.5}$ & 11.11$^{+0.21}_{-0.05}$ & 0.3$^{+0.5}_{-0.1}$ & 4.3$^{+0.2}_{-3.1}$ & 11.19$^{+0.07}_{-0.42}$ & 0.5$^{+0.3}_{-0.4}$ & 2.3$^{+2.2}_{-1.2}$ & 10.94$^{+0.22}_{-0.41}$ & 0.3$^{+0.5}_{-0.1}$ \\
5634 & 4.0$^{+0.3}_{-2.2}$ & 11.40$^{+0.02}_{-0.16}$ & 0.8$^{+0.3}_{-0.7}$ & 1.9$^{+2.8}_{-0.5}$ & 11.08$^{+0.29}_{-0.15}$ & 0.1$^{+0.7}_{-0.0}$ & 1.8$^{+2.5}_{-0.7}$ & 11.01$^{+0.33}_{-0.24}$ & 0.3$^{+0.5}_{-0.1}$ \\
5693 & - & - & - & - & - & - & - & - & - \\
5680 & - & - & - & - & - & - & - & - & - \\
5794 & - & - & - & - & - & - & - & - & - \\
8495 & - & - & - & - & - & - & - & - & - \\
5748 & 4.4$^{+0.3}_{-1.3}$ & 11.44$^{+0.05}_{-0.18}$ & 0.1$^{+0.4}_{-0.0}$ & 4.7$^{+0.2}_{-1.3}$ & 11.42$^{+0.03}_{-0.15}$ & 0.1$^{+0.4}_{-0.0}$ & 3.3$^{+1.0}_{-1.3}$ & 11.26$^{+0.18}_{-0.10}$ & 0.5$^{+0.3}_{-0.4}$ \\
5689 & 1.9$^{+2.0}_{-0.3}$ & 10.68$^{+0.16}_{-0.09}$ & 0.1$^{+0.7}_{-0.0}$ & 3.3$^{+0.8}_{-2.8}$ & 10.76$^{+0.09}_{-0.45}$ & 0.8$^{+0.3}_{-0.7}$ & 0.9$^{+2.5}_{-0.1}$ & 10.33$^{+0.37}_{-0.10}$ & 0.1$^{+0.7}_{-0.0}$ \\
5876 & 3.8$^{+0.3}_{-1.3}$ & 11.42$^{+0.07}_{-0.09}$ & 0.8$^{+0.3}_{-0.4}$ & 4.1$^{+0.2}_{-3.3}$ & 11.41$^{+0.03}_{-0.41}$ & 0.8$^{+0.3}_{-0.7}$ & 3.5$^{+0.3}_{-2.2}$ & 11.35$^{+0.06}_{-0.29}$ & 0.8$^{+0.3}_{-0.5}$ \\
5602 & 2.1$^{+0.8}_{-0.2}$ & 10.98$^{+0.12}_{-0.07}$ & 0.1$^{+0.3}_{-0.0}$ & 2.9$^{+1.8}_{-1.5}$ & 11.05$^{+0.14}_{-0.36}$ & 0.1$^{+0.7}_{-0.0}$ & 1.8$^{+1.0}_{-0.6}$ & 10.80$^{+0.18}_{-0.24}$ & 0.3$^{+0.3}_{-0.1}$ \\
8662 & 3.8$^{+0.3}_{-1.0}$ & 10.96$^{+0.03}_{-0.11}$ & 1.0$^{+0.5}_{-0.3}$ & 3.8$^{+0.2}_{-3.8}$ & 10.93$^{+0.03}_{-0.56}$ & 1.0$^{+0.5}_{-0.9}$ & 3.6$^{+0.3}_{-3.1}$ & 10.85$^{+0.08}_{-0.46}$ & 1.0$^{+0.5}_{-0.8}$ \\
8041 & 1.2$^{+1.2}_{-0.3}$ & 10.76$^{+0.26}_{-0.04}$ & 0.1$^{+0.3}_{-0.0}$ & 0.8$^{+0.6}_{-0.2}$ & 10.60$^{+0.17}_{-0.15}$ & 0.1$^{+0.1}_{-0.0}$ & 0.9$^{+0.3}_{-0.1}$ & 10.64$^{+0.03}_{-0.07}$ & 0.1$^{+0.1}_{-0.0}$ \\
8625 & 1.2$^{+0.6}_{-0.4}$ & 10.43$^{+0.12}_{-0.05}$ & 0.3$^{+0.1}_{-0.1}$ & 0.6$^{+0.5}_{-0.1}$ & 10.13$^{+0.11}_{-0.08}$ & 0.1$^{+0.1}_{-0.0}$ & 0.7$^{+0.2}_{-0.1}$ & 10.23$^{+0.10}_{-0.06}$ & 0.1$^{+0.1}_{-0.0}$ \\
7653 & 2.8$^{+1.5}_{-1.3}$ & 10.80$^{+0.13}_{-0.12}$ & 0.5$^{+0.5}_{-0.4}$ & 1.8$^{+2.8}_{-1.0}$ & 10.59$^{+0.31}_{-0.30}$ & 0.3$^{+0.8}_{-0.1}$ & 1.2$^{+3.3}_{-0.2}$ & 10.34$^{+0.52}_{-0.06}$ & 0.3$^{+0.8}_{-0.0}$ \\
8047 & 1.5$^{+0.8}_{-0.3}$ & 10.86$^{+0.09}_{-0.15}$ & 0.3$^{+0.3}_{-0.0}$ & 0.6$^{+0.1}_{-0.1}$ & 10.47$^{+0.02}_{-0.04}$ & 0.1$^{+0.1}_{-0.0}$ & 0.8$^{+0.1}_{-0.1}$ & 10.60$^{+0.08}_{-0.07}$ & 0.1$^{+0.1}_{-0.0}$ \\
7475 & 2.3$^{+2.2}_{-0.5}$ & 11.09$^{+0.22}_{-0.15}$ & 0.3$^{+0.8}_{-0.1}$ & 1.1$^{+3.6}_{-0.2}$ & 10.72$^{+0.46}_{-0.09}$ & 0.1$^{+0.9}_{-0.0}$ & 1.5$^{+1.3}_{-0.4}$ & 10.81$^{+0.26}_{-0.15}$ & 0.3$^{+0.3}_{-0.1}$ \\
\cutinhead{\textsc{Lynx Cluster W ($\langle z \rangle = 1.273$)}}
1745 & 2.6$^{+1.8}_{-1.0}$ & 10.78$^{+0.16}_{-0.10}$ & 0.4$^{+0.4}_{-0.3}$ & 3.0$^{+1.3}_{-2.5}$ & 10.76$^{+0.13}_{-0.49}$ & 0.5$^{+0.3}_{-0.4}$ & 1.5$^{+3.0}_{-0.3}$ & 10.43$^{+0.40}_{-0.14}$ & 0.3$^{+0.8}_{-0.1}$ \\
1486 & 3.6$^{+0.3}_{-2.0}$ & 10.79$^{+0.06}_{-0.19}$ & 1.0$^{+0.5}_{-0.5}$ & 3.3$^{+0.6}_{-2.2}$ & 10.70$^{+0.11}_{-0.28}$ & 1.0$^{+0.5}_{-0.6}$ & 2.3$^{+1.8}_{-1.7}$ & 10.54$^{+0.18}_{-0.34}$ & 0.8$^{+0.8}_{-0.5}$ \\
1794 & 3.3$^{+1.0}_{-1.6}$ & 11.03$^{+0.08}_{-0.20}$ & 0.5$^{+0.3}_{-0.4}$ & 2.3$^{+2.3}_{-1.0}$ & 10.80$^{+0.25}_{-0.21}$ & 0.3$^{+0.5}_{-0.1}$ & 1.2$^{+3.5}_{-0.3}$ & 10.51$^{+0.60}_{-0.03}$ & 0.1$^{+0.7}_{-0.0}$ \\
1922 & 1.9$^{+1.3}_{-0.3}$ & 11.14$^{+0.12}_{-0.02}$ & 0.1$^{+0.4}_{-0.0}$ & 1.8$^{+2.5}_{-0.8}$ & 11.03$^{+0.29}_{-0.23}$ & 0.3$^{+0.5}_{-0.1}$ & 1.5$^{+1.3}_{-0.3}$ & 10.91$^{+0.26}_{-0.04}$ & 0.3$^{+0.3}_{-0.1}$ \\
1525 & 2.6$^{+1.8}_{-1.0}$ & 11.04$^{+0.17}_{-0.10}$ & 0.4$^{+0.4}_{-0.3}$ & 2.0$^{+2.6}_{-1.0}$ & 10.85$^{+0.31}_{-0.26}$ & 0.3$^{+0.5}_{-0.1}$ & 1.5$^{+2.6}_{-0.3}$ & 10.69$^{+0.38}_{-0.15}$ & 0.3$^{+0.5}_{-0.1}$ \\
1962 & 3.0$^{+1.3}_{-1.3}$ & 10.97$^{+0.12}_{-0.12}$ & 0.5$^{+0.3}_{-0.4}$ & 2.1$^{+2.6}_{-0.7}$ & 10.79$^{+0.26}_{-0.18}$ & 0.1$^{+0.7}_{-0.0}$ & 1.8$^{+2.2}_{-0.3}$ & 10.69$^{+0.26}_{-0.14}$ & 0.3$^{+0.5}_{-0.1}$ \\
2094 & 4.0$^{+0.3}_{-1.8}$ & 11.21$^{+0.13}_{-0.11}$ & 0.8$^{+0.3}_{-0.7}$ & 4.1$^{+0.2}_{-3.1}$ & 11.17$^{+0.14}_{-0.37}$ & 0.8$^{+0.3}_{-0.7}$ & 3.8$^{+0.3}_{-3.2}$ & 11.16$^{+0.06}_{-0.59}$ & 0.8$^{+0.3}_{-0.7}$ \\
2343 & 3.8$^{+0.5}_{-1.3}$ & 11.65$^{+0.11}_{-0.13}$ & 0.5$^{+0.3}_{-0.4}$ & 4.0$^{+0.3}_{-2.5}$ & 11.62$^{+0.10}_{-0.31}$ & 0.5$^{+0.3}_{-0.4}$ & 2.5$^{+2.0}_{-0.8}$ & 11.46$^{+0.18}_{-0.15}$ & 0.3$^{+0.5}_{-0.1}$ \\
2195 & 4.0$^{+0.3}_{-1.5}$ & 11.54$^{+0.03}_{-0.12}$ & 0.8$^{+0.3}_{-0.4}$ & 2.0$^{+2.6}_{-0.7}$ & 11.21$^{+0.31}_{-0.13}$ & 0.3$^{+0.5}_{-0.1}$ & 2.4$^{+1.8}_{-1.5}$ & 11.31$^{+0.19}_{-0.37}$ & 0.4$^{+0.4}_{-0.3}$ \\
2571 & 2.8$^{+1.8}_{-0.5}$ & 11.14$^{+0.17}_{-0.07}$ & 0.3$^{+0.5}_{-0.1}$ & 3.4$^{+1.0}_{-1.8}$ & 11.15$^{+0.15}_{-0.26}$ & 0.4$^{+0.4}_{-0.3}$ & 1.9$^{+2.8}_{-0.3}$ & 10.91$^{+0.32}_{-0.11}$ & 0.1$^{+0.7}_{-0.0}$ \\
\enddata
\tablecomments{Ages ($\langle t \rangle_{SFW}$), stellar masses and SFH $\tau$ are derived by fitting galaxy SEDs with BC03/MA05/CB07 models. For the photometry, we use 1.5" radius aperture photometry and aperture corrections are estimated with PSF growth curves normalized at 7" radius.
If one wants to use the stellar masses derived with photometry using simulated galaxies growth curve for aperture correction, see \textsc{Tab}.\ref{tab:fit_param_influence}.
The SFH $\tau$ are given as a reference for age estimation: these values are very uncertain as stated in the text.}
\end{deluxetable}

\begin{deluxetable}{l c c c || c c c || c c c}
	\setlength{\tabcolsep}{0.15in}
	\tabletypesize{\scriptsize}
%	\rotate
%	\tablewidth{0pt}
%	\tablenum{}
	\tablecolumns{7}
	\tablecaption{Lynx group ETG stellar population ages and stellar masses \label{tab:cat_mass_gr}}
	\tablehead{
		\colhead{} 	& 
		\multicolumn{3}{c}{BC03} & 
		\multicolumn{3}{c}{MA05} & 
		\multicolumn{3}{c}{CB07} \\
		\multicolumn{10}{c}{} \\
		\colhead{ID} & 
		\colhead{Age} & 
		\colhead{log $\frac{M}{M_{\sun}}$} & 
		\colhead{SFH $\tau$} & 
		\colhead{Age} & 
		\colhead{log $\frac{M}{M_{\sun}}$} & 
		\colhead{SFH $\tau$} & 
		\colhead{Age} & 
		\colhead{log $\frac{M}{M_{\sun}}$} & 
		\colhead{SFH $\tau$} \\
		\colhead{} & 
		\colhead{(Gyrs)} & 
		\colhead{} & 
		\colhead{(Gyrs)} & 
		\colhead{(Gyrs)} & 
		\colhead{} & 
		\colhead{(Gyrs)} & 
		\colhead{(Gyrs)} &
		\colhead{} &
		\colhead{(Gyrs)}
	}
	\startdata
\cutinhead{\textsc{Lynx Group 1 ($\langle z \rangle = 1.266$)}}
518 & 1.3$^{+3.0}_{-1.0}$ & 10.40$^{+0.28}_{-0.16}$ & 0.5$^{+2.5}_{-0.4}$ & 0.4$^{+4.3}_{-0.1}$ & 10.05$^{+0.54}_{-0.07}$ & 0.1$^{+4.9}_{-0.0}$ & 0.7$^{+0.4}_{-0.4}$ & 10.16$^{+0.10}_{-0.16}$ & 0.3$^{+0.3}_{-0.1}$ \\
1339 & 3.5$^{+0.8}_{-1.0}$ & 11.59$^{+0.14}_{-0.10}$ & 0.5$^{+0.3}_{-0.4}$ & 4.0$^{+0.3}_{-2.5}$ & 11.59$^{+0.10}_{-0.31}$ & 0.5$^{+0.3}_{-0.4}$ & 2.6$^{+1.8}_{-1.7}$ & 11.43$^{+0.19}_{-0.42}$ & 0.4$^{+0.4}_{-0.3}$ \\
1024 & 1.2$^{+0.6}_{-0.2}$ & 10.64$^{+0.13}_{-0.03}$ & 0.3$^{+0.1}_{-0.0}$ & 0.6$^{+0.1}_{-0.1}$ & 10.36$^{+0.03}_{-0.08}$ & 0.1$^{+0.1}_{-0.0}$ & 0.8$^{+0.1}_{-0.1}$ & 10.50$^{+0.03}_{-0.08}$ & 0.1$^{+0.1}_{-0.0}$ \\
825 & 2.1$^{+1.3}_{-1.2}$ & 11.41$^{+0.11}_{-0.24}$ & 0.4$^{+0.4}_{-0.3}$ & 0.8$^{+2.6}_{-0.1}$ & 11.01$^{+0.43}_{-0.07}$ & 0.1$^{+0.7}_{-0.0}$ & 0.9$^{+1.0}_{-0.1}$ & 11.06$^{+0.15}_{-0.06}$ & 0.1$^{+0.3}_{-0.0}$ \\
1249 & - & - & - & - & - & - & - & - & - \\
1085 & 2.1$^{+2.6}_{-0.2}$ & 11.28$^{+0.16}_{-0.07}$ & 0.1$^{+0.9}_{-0.0}$ & 1.4$^{+3.3}_{-0.3}$ & 11.07$^{+0.31}_{-0.06}$ & 0.1$^{+0.9}_{-0.0}$ & 1.8$^{+2.8}_{-0.6}$ & 11.15$^{+0.28}_{-0.24}$ & 0.3$^{+0.8}_{-0.1}$ \\
\multicolumn{10}{c}{} \\
\cutinhead{\textsc{Lynx Group 2 ($\langle z \rangle = 1.262$)}}
\multicolumn{10}{c}{} \\
1636 & 2.5$^{+2.0}_{-0.6}$ & 11.24$^{+0.16}_{-0.08}$ & 0.3$^{+0.5}_{-0.1}$ & 2.3$^{+2.3}_{-1.3}$ & 11.11$^{+0.24}_{-0.33}$ & 0.3$^{+0.5}_{-0.1}$ & 1.8$^{+2.0}_{-0.6}$ & 11.01$^{+0.23}_{-0.25}$ & 0.3$^{+0.5}_{-0.1}$ \\
939 & 1.2$^{+0.6}_{-0.4}$ & 10.99$^{+0.12}_{-0.03}$ & 0.3$^{+0.1}_{-0.1}$ & 0.6$^{+0.1}_{-0.1}$ & 10.71$^{+0.03}_{-0.10}$ & 0.1$^{+0.1}_{-0.0}$ & 0.7$^{+0.1}_{-0.1}$ & 10.79$^{+0.08}_{-0.03}$ & 0.1$^{+0.1}_{-0.0}$ \\
1383 & 3.0$^{+1.3}_{-1.3}$ & 11.18$^{+0.14}_{-0.15}$ & 0.5$^{+0.3}_{-0.4}$ & 1.4$^{+3.3}_{-0.5}$ & 10.84$^{+0.42}_{-0.25}$ & 0.1$^{+0.7}_{-0.0}$ & 1.5$^{+1.3}_{-0.4}$ & 10.79$^{+0.26}_{-0.14}$ & 0.3$^{+0.3}_{-0.1}$ \\
2000 & 4.0$^{+0.3}_{-1.8}$ & 11.61$^{+0.03}_{-0.12}$ & 0.8$^{+0.3}_{-0.7}$ & 4.1$^{+0.2}_{-3.3}$ & 11.57$^{+0.07}_{-0.41}$ & 0.8$^{+0.3}_{-0.7}$ & 3.5$^{+0.5}_{-3.0}$ & 11.51$^{+0.10}_{-0.51}$ & 0.8$^{+0.3}_{-0.7}$ \\
2519 & 3.3$^{+0.8}_{-1.8}$ & 10.97$^{+0.06}_{-0.15}$ & 0.8$^{+0.3}_{-0.7}$ & 0.9$^{+3.8}_{-0.1}$ & 10.46$^{+0.55}_{-0.08}$ & 0.1$^{+0.9}_{-0.0}$ & 1.2$^{+2.8}_{-0.2}$ & 10.47$^{+0.41}_{-0.03}$ & 0.3$^{+0.8}_{-0.0}$ \\
1791 & 1.2$^{+0.8}_{-0.4}$ & 11.15$^{+0.14}_{-0.07}$ & 0.3$^{+0.3}_{-0.1}$ & 0.5$^{+0.1}_{-0.1}$ & 10.81$^{+0.09}_{-0.03}$ & 0.1$^{+0.1}_{-0.0}$ & 0.7$^{+0.1}_{-0.1}$ & 10.97$^{+0.07}_{-0.08}$ & 0.1$^{+0.1}_{-0.0}$ \\
\cutinhead{\textsc{Lynx Group 3 ($\langle z \rangle = 1.264$)}}
137 & 3.1$^{+1.5}_{-0.5}$ & 11.20$^{+0.19}_{-0.06}$ & 0.1$^{+0.7}_{-0.0}$ & 0.9$^{+1.5}_{-0.1}$ & 10.65$^{+0.42}_{-0.04}$ & 0.1$^{+0.1}_{-0.0}$ & 1.3$^{+2.1}_{-0.2}$ & 10.73$^{+0.40}_{-0.12}$ & 0.1$^{+0.4}_{-0.0}$ \\
542 & 2.7$^{+2.0}_{-0.6}$ & 11.34$^{+0.15}_{-0.12}$ & 0.1$^{+0.7}_{-0.0}$ & 1.6$^{+3.1}_{-0.2}$ & 11.06$^{+0.36}_{-0.06}$ & 0.1$^{+0.7}_{-0.0}$ & 2.1$^{+2.2}_{-1.1}$ & 11.14$^{+0.23}_{-0.31}$ & 0.4$^{+0.6}_{-0.3}$ \\
1135 & 2.6$^{+1.8}_{-1.0}$ & 11.32$^{+0.13}_{-0.10}$ & 0.4$^{+0.6}_{-0.3}$ & 1.8$^{+2.8}_{-1.0}$ & 11.11$^{+0.28}_{-0.29}$ & 0.3$^{+0.8}_{-0.1}$ & 1.8$^{+2.6}_{-0.8}$ & 11.07$^{+0.30}_{-0.23}$ & 0.4$^{+0.6}_{-0.1}$ \\
889 & 3.1$^{+1.5}_{-0.5}$ & 10.90$^{+0.16}_{-0.07}$ & 0.1$^{+0.7}_{-0.0}$ & 4.4$^{+0.2}_{-2.8}$ & 10.96$^{+0.06}_{-0.34}$ & 0.4$^{+0.4}_{-0.3}$ & 1.2$^{+3.5}_{-0.3}$ & 10.35$^{+0.60}_{-0.03}$ & 0.1$^{+0.9}_{-0.0}$ \\
1431 & 2.4$^{+2.2}_{-0.3}$ & 11.01$^{+0.15}_{-0.07}$ & 0.1$^{+0.9}_{-0.0}$ & 1.8$^{+2.8}_{-1.0}$ & 10.80$^{+0.33}_{-0.29}$ & 0.3$^{+0.8}_{-0.1}$ & 2.0$^{+2.2}_{-1.1}$ & 10.79$^{+0.27}_{-0.27}$ & 0.5$^{+0.5}_{-0.4}$ \\
1064 & 1.5$^{+2.8}_{-0.7}$ & 10.53$^{+0.20}_{-0.14}$ & 0.5$^{+1.5}_{-0.3}$ & 1.7$^{+2.3}_{-1.6}$ & 10.37$^{+0.22}_{-0.26}$ & 1.0$^{+2.0}_{-0.8}$ & 0.8$^{+1.7}_{-0.1}$ & 10.22$^{+0.20}_{-0.04}$ & 0.3$^{+1.3}_{-0.0}$ \\
1136 & 3.1$^{+1.3}_{-0.8}$ & 11.29$^{+0.18}_{-0.05}$ & 0.4$^{+0.4}_{-0.3}$ & 4.0$^{+0.3}_{-2.8}$ & 11.33$^{+0.10}_{-0.36}$ & 0.5$^{+0.3}_{-0.4}$ & 1.9$^{+2.8}_{-0.7}$ & 11.04$^{+0.30}_{-0.31}$ & 0.1$^{+0.9}_{-0.0}$ \\
1775 & 3.8$^{+0.3}_{-2.6}$ & 11.19$^{+0.02}_{-0.17}$ & 1.0$^{+0.5}_{-0.9}$ & 2.1$^{+2.3}_{-1.5}$ & 10.93$^{+0.26}_{-0.32}$ & 0.4$^{+0.6}_{-0.3}$ & 2.5$^{+1.5}_{-1.8}$ & 10.97$^{+0.19}_{-0.34}$ & 0.8$^{+0.3}_{-0.5}$ \\
1731 & 1.7$^{+2.6}_{-0.8}$ & 11.12$^{+0.21}_{-0.11}$ & 0.5$^{+1.0}_{-0.3}$ & 2.6$^{+1.0}_{-2.8}$ & 11.08$^{+0.09}_{-0.39}$ & 1.5$^{+0.5}_{-1.3}$ & 0.8$^{+1.3}_{-0.1}$ & 10.86$^{+0.10}_{-0.11}$ & 0.1$^{+0.7}_{-0.0}$\\
\enddata
\tablecomments{Cf. \textsc{Tab}.\ref{tab:cat_mass_cl}}
\end{deluxetable}

%@@@@@@@@@@@@@@@@@@@@@@@@@@@@@@@@@@@@@@@@@@@@@@@@@@@@@@@@@@@@@@@@@@@@@@@@
% BIBLIOGRAPHY
%@@@@@@@@@@@@@@@@@@@@@@@@@@@@@@@@@@@@@@@@@@@@@@@@@@@@@@@@@@@@@@@@@@@@@@@@

\clearpage
%\newpage

\end{document}